\documentclass[12pt,preprint]{aastex}

\shorttitle{M37 Transit Survey I}
\shortauthors{Hartman et al.}

\begin{document}

\citeindextrue

\title{Deep MMT\altaffilmark{1}
Transit Survey of the Open Cluster M37 I: Observations and Cluster Parameters}
\author{J.~D.~Hartman\altaffilmark{2}, B.~S.~Gaudi\altaffilmark{3}, M.~J.~Holman\altaffilmark{2}, B.~A.~McLeod\altaffilmark{2}, K.~Z.~Stanek\altaffilmark{3}, J.~A.~Barranco\altaffilmark{4}, M.~H.~Pinsonneault\altaffilmark{3}, S.~Meibom\altaffilmark{2}, and J.~S.~Kalirai\altaffilmark{5,6}}

\altaffiltext{1}{Observations reported here were obtained at the MMT Observatory, a joint facility of the Smithsonian Institution and the University of Arizona.}
\altaffiltext{2}{Harvard-Smithsonian Center for Astrophysics, 60 Garden St., Cambridge, MA~02138, USA; jhartman@cfa.harvard.edu, mholman@cfa.harvard.edu, bmcleod@cfa.harvard.edu, smeibom@cfa.harvard.edu}
\altaffiltext{3}{Department of Astronomy, The Ohio State University, Columbus, OH~43210, USA; gaudi@astronomy.ohio-state.edu, kstanek@astronomy.ohio-state.edu, pinsono@astronomy.ohio-state.edu}
\altaffiltext{4}{Department of Physics and Astronomy, San Francisco State University, 1600 Holloway Ave., San Francisco, CA~94132, USA; barranco@stars.sfsu.edu}
\altaffiltext{5}{University of California Observatories/Lick Observatory, University of California at Santa Cruz, Santa Cruz CA, 95060; jkalirai@ucolick.org}
\altaffiltext{6}{Hubble Fellow}

\begin{abstract}
We have conducted a deep ($15 \la r \la 23$), 20 night survey for transiting planets in the intermediate age open cluster M37 (NGC 2099) using the Megacam wide-field mosaic CCD camera on the 6.5m Multiple Mirror Telescope (MMT). In this paper we describe the observations and data reduction procedures for the survey and analyze the stellar content and dynamical state of the cluster. By combining high resolution spectroscopy with existing $BVI_{C}K_{S}$ and new $gri$ color magnitude diagrams we determine the fundamental cluster parameters: $t = 485 \pm 28$ Myr without overshooting ($t = 550 \pm 30~{\rm Myr}$ with overshooting), $E(B-V) = 0.227 \pm 0.038$, $(m-M)_{V} = 11.57 \pm 0.13$ and $[M/H] = +0.045 \pm 0.044$ which are in good agreement with, though more precise than, previous measurements. We determine the mass function down to $0.3 M_{\odot}$ and use this to estimate the total cluster mass of $3640 \pm 170 M_{\odot}$. 
\end{abstract}

\keywords{open clusters and associations:individual (M37) --- surveys --- stars: luminosity function, mass function --- stars:fundamental parameters --- planetary systems}

\section{Introduction}

In recent years the discovery and characterization of exoplanets that transit their host stars has revolutionized our understanding of planetary systems \citep[See][for a review]{Charbonneau.06}. The search for these planets has proceeded at a frantic pace leading to a proliferation of surveys following different strategies to find these elusive systems. A number of groups have opted to search for planets in galactic open clusters. This includes the UStAPS \citep{Street.03,Bramich.05,Hood.05}, EXPLORE-OC \citep{vonBraun.05}, PISCES \citep{Mochejska.05, Mochejska.06}, STEPSS \citep{Burke.06} and MONITOR \citep{Aigrain.07} projects as well as a survey by \citet{Montalto.07}.  There has also been significant work invested in developing optimum strategies to search for planets in open clusters \citep{Janes.96,vonBraun.05,Pepper.05}. The rationale behind searching for transiting planets in open clusters is that since the properties of stars in clusters are easier to determine en masse than they are for field stars, and since cluster stars have more uniform properties than field stars, one can determine in a relatively straightforward fashion from the survey the frequency of planets in the cluster, or at least place a meaningful upper limit on it \citep[e.g.][]{Burke.06}.

While to date no confirmed transiting planet has been found in an open cluster, these surveys have produced a number of by-products unrelated to the study of planets. This includes improving the fundamental parameters (age, distance, reddening and metallicity) of clusters \citep[e.g.][]{Burke.04}, studying the general population of variable stars found in and around the clusters \citep{Mochejska.02, Mochejska.04,Pepper.06b,DeMarchi.07}, studying low mass stellar rotation \citep{Irwin.06,Irwin.07a}, and the discovery of a very low mass pre-main sequence eclipsing binary system \citep{Irwin.07b}.

In this paper we introduce a deep transit survey of the intermediate age ($\sim 500~{\rm Myr}$) open cluster M37 (NGC 2099) using the MMT. The survey was conducted over twenty nights between Dec. 2005 and Jan 2006, and we accumulated over 4000 quality images of the cluster. This is easily the largest telescope ever utilized for such a survey, and as a by-product we have conducted a very deep survey for variable stars in this cluster (down to $r \sim 23$). 

We were motivated to conduct this transit survey by \citet{Pepper.05,Pepper.06a} who suggested that it may be possible to find Neptune-sized planets transiting solar-like stars by surveying an open cluster with a large telescope. Preliminary observations of the well-studied open cluster NGC 6791 with the MMT suggested that this was indeed technically feasible \citep{Hartman.05}. 

The results of the survey related to variable stars, transiting planets and stellar rotation will be presented in future papers, our goal in this paper is to present the observations, describe the data reduction procedures, and study the stars in our field. This includes refining estimates for the fundamental cluster parameters and studying the distribution of stars (mass/luminosity functions) and the dynamic state of the cluster. Having good estimates for these parameters is essential for all subsequent results from this survey. In particular, it is necessary to know these parameters in order to obtain masses and radii for the cluster stars as a function of magnitude which in turn is needed to determine the frequency of planets from our transit survey.

This paper is certainly not the first determination of the fundamental parameters for M37. Since 1996 there have been at least seven published determinations of its age, distance and reddening \citep{Mermilliod.96,Kiss.01,Kalirai.01,Nilakshi.02,Sarajedini.04,Kalirai.05,Kang.07}, a spectroscopic determination of its metallicity \citep{Marshall.05}, three photometric estimates of its metallicity \citep{Mermilliod.96, Twarog.97, Nilakshi.02}, and two surveys for variable stars \citep{Kiss.01,Kang.07}. The age estimates vary between $400-650$ Myr, the distance modulus is $(m-M)_{V} \sim 11.5$, the reddening estimates vary between $E(B-V) = 0.21-0.3$, and the metallicity estimates between $[Fe/H] = -0.2$ and $+0.1$. As we will show below, our determination of the fundamental parameters is in good agreement with the previous values, but we do improve the precision of these measurements by using multi-color photometry, high resolution spectroscopy for a number of cluster members, new empirically calibrated isochrones due to \citet{An.07} and a detailed error analysis. 

In the following section we will describe our observations which consist of $r$-band photometric time series data, $gri$ photometry, and high resolution spectroscopy all obtained at the MMT. We will describe the photometric data reduction in \S 3 and present spectroscopic results in \S 4. Using the reduced spectra we will determine the temperatures, radial velocities and estimated metallicities for 127 stars in our field. We will then present and analyze color-magnitude diagrams in \S 5, which when combined with the spectroscopic estimates for the metallicity and reddening of the cluster will allow us to obtain new estimates for the cluster fundamental parameters. In \S 6 we present the mass and luminosity functions of the cluster and in \S 7 we discuss the radial density profile of the cluster. We conclude in \S 8.

\section{Observations}

\subsection{Photometric Observations}

The photometric observations were obtained over twenty four nights (eight of which were half nights) between December 21, 2005 and January 21, 2006, as summarized in table~\ref{tab:obssummary}. We obtained a total of $\sim$ 5000 images using the Megacam instrument \citep{McLeod.00} on the MMT. Megacam is a $24\arcmin \times 24\arcmin$ mosaic imager consisting of 36 2k$\times$4k, thinned, backside-illuminated CCDs that are each read out by two amplifiers. The mosaic has an unbinned pixel scale of $0.08\arcsec$ which allows for a well sampled point-spread-function (PSF) even under the best seeing conditions. To decrease the read-out time we used $2 \times 2$ binning with the gain set so that the pixel sensitivity became non-linear before the analog-to-digital conversion threshold of 65536 counts. Because of the fine sampling and the relatively deep pixel wells, one can collect $2\times 10^7$ photons in $1\arcsec$ seeing from a single star prior to saturation, setting the photon limit on the precision in a single exposure to $\sim$ 0.25 mmag.

\placetable{tab:obssummary}

\begin{figure}[p]
\plotone{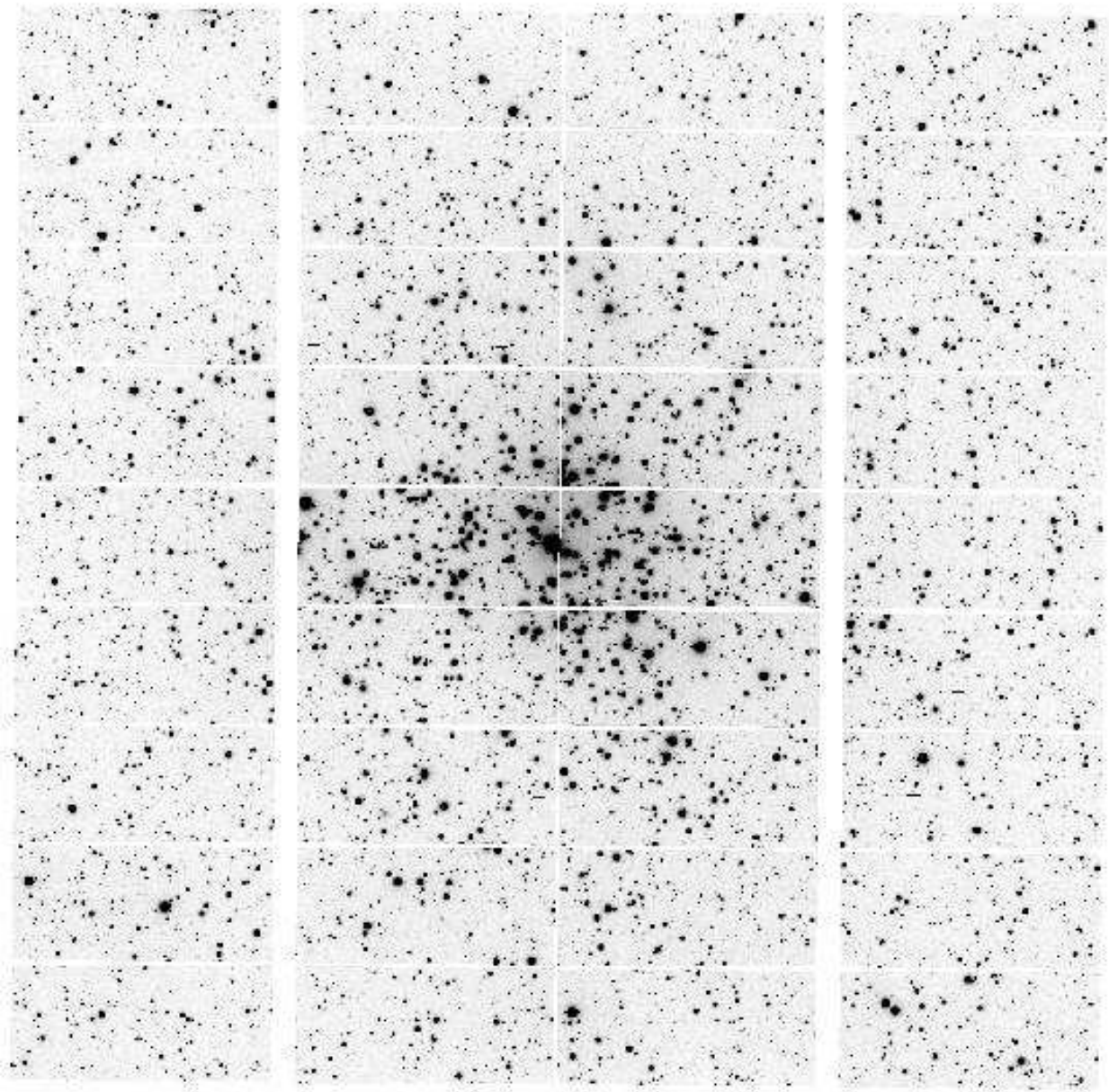}
\caption{A typical $r^{\prime}$ Megacam mosaic image of M37. The field of view is $24\arcmin$x$24\arcmin$, north is up and east is to the left. The numbering convention for the chips used in the text starts in the upper left corner and increases going down.}
\label{fig:M37_fov}
\end{figure}

The time series observations were centered on 05:52:19, +32:33:12 (J2000). We used the same guide stars for the entire run so that the pointing was stable to within a few pixels, this mitigates the effect of errors in the flat-field. We used an $r^{\prime}$ filter to maximize the sensitivity to smaller stars while avoiding fringing that would occur at longer wavelengths. The average time between exposures (including read-out and time spent initializing for the next exposure) was 24 seconds. Figure~\ref{fig:M37_fov} shows a typical mosaic image of the cluster.

\placefigure{fig:M37_fov}

Ideally one would like to observe the same stars from exposure to exposure while spending as little time as necessary with the shutter closed. In other words, one would like to maximize the exposure times on a fixed set of stars. As the exposure time is increased more stars are lost to saturation, and, more importantly, the fraction of the image that is lost to saturated stars and artifacts (diffraction spikes and bleed columns) increases. Not only does this decrease the number of stars that can be observed, it also compromises the quality of the image and the ability to reduce the image to achieve high precision photometry for any of the stars. To determine the optimal exposure time we obtained a series of preliminary observations of M37 during a Megacam engineering run on October 29, 2005. We found that in $1\arcsec$ seeing an exposure time of sixty seconds was the longest we could go before saturation posed a problem for image quality. 

The exposure time of each image was chosen to keep a $r \sim 15$ mag star as close to the saturation limit as possible while not allowing more than $2\%$ of the stars located near the main sequence on a color magnitude diagram (CMD) to be contaminated by artifacts due to other saturated stars. After reading out an image we would calculate a suggested exposure time for the next image based on the seeing and atmospheric transparency. In calculating the suggested exposure time we used the Source Extractor program (version 2.3.2) due to \citet{Bertin.96} to measure the FWHM and peak pixel value of a fiducial star after each image was downloaded from the camera. We then calculated the exposure time that would be necessary to keep the peak pixel value of that star at the desired level:
\begin{equation}
s_{1} = EXPTIME \times \frac{p_0}{p}
\end{equation}
where $EXPTIME$ is the previous exposure time, $p$ is the measured peak pixel value, and $p_{0}$ is the desired peak pixel value. We also calculated the exposure time that would be necessary to keep constant the fraction of stars unaffected by artifacts that scale with the percentage of the image that is saturated using the following empirical relation:
\begin{equation}
s_{2} = \frac{a_{1}}{(FWHM/{\rm pixels})^{a_2}} \frac{EXPTIME}{(p/ADU)}
\end{equation}
where $a_1$ and $a_2$ are parameters that we determined empirically for our field ($a_1 \sim 5 \times 10^{5}$ and $a_2 \sim 1.3$). The exposure time is then given as the minimum of $s_1$, $s_2$ and $30$ seconds, where we adopt a lower bound of $30$ seconds to avoid spending more than half the time with the shutter closed. Typical exposure times as a function of seeing for photometric conditions are listed in table~\ref{tab:exptimes}.

\placetable{tab:exptimes}

In addition to the time series observations, we also obtained a number of observations to use in constructing color-magnitude diagrams of the cluster. These include 40 $g^{\prime}$ and 35 $i^{\prime}$ observations of our primary field obtained over four nights (12/23, 12/24, 12/27 and 12/30). To observe the cluster main sequence turn-off and red giant stars we took a set of short exposure images ($1-2$ seconds) as well as longer exposure images to match the $r^{\prime}$ time series images. To calibrate our data we observed two fields covered by the Sloan Digital Sky Survey Data Release 5 \citep[SDSS;][]{Adelman-McCarthy.07}. The first of these fields, located at 03:20:00, 00:00:00 (J2000) was observed during photometric conditions on 12/24 and 12/27. The second field, located at 08:00:00, +35:00:00 (J2000) was observed during brief photometric conditions on 12/28, 12/30 and 01/05. The $i^{\prime}$ observations of the second field were only obtained on 12/30.

Finally, to estimate the cluster membership contamination we observed a field centered at 05:57:02 +30:49:32 (J2000). The field is located two degrees from the primary M37 field and at the same galactic latitude. The observations for this field were obtained on 12/30 and 01/07 using an exposure time of $60-70$ seconds.

\subsection{Spectroscopic Observations}

\begin{figure}[p]
\plotone{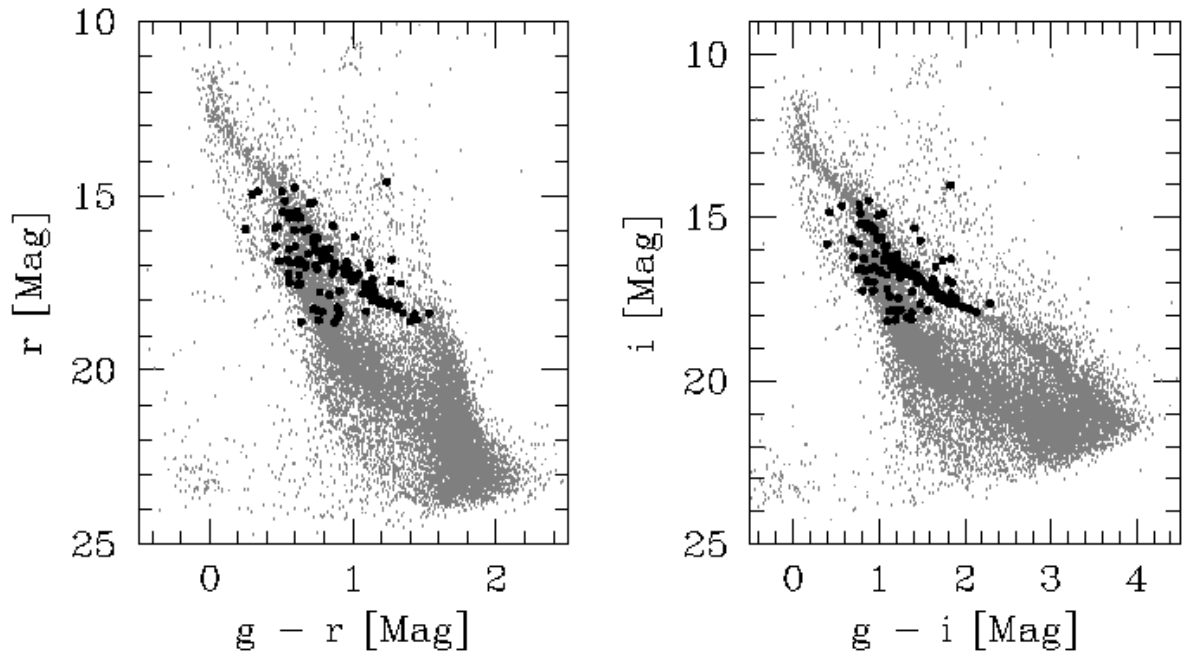}
\caption{The targets brighter than $r = 18.65$ for which high resolution spectroscopy was obtained are plotted as dark points on $g-r$ and $g-i$ CMDs of the cluster.}
\label{fig:spectargetsonCMD}
\end{figure}

To follow-up our photometric observations we obtained spectra for 167 stars in the field of M37 with $14.5 < r < 22.2$. For stars with $r > 18.65$ the signal to noise was too low to be useful for spectroscopic classification/radial velocity measurements. In figure~\ref{fig:spectargetsonCMD} we show the 127 spectroscopic targets with $r < 18.65$ on $g-r$ and $g-i$ CMDs. The targets include a number of candidate transiting planets, eclipsing binaries and other large-amplitude variables which we will describe elsewhere. There are 74 bright targets that lie near the cluster main sequence in $g-r$ and $g-i$ CMDs and will be used to determine the reddening, metallicity, and radial velocity of the cluster. The latter quantity is used to assess the membership probability for transit candidates that are likely members based on their photometry. Other targets not near the main sequence are used as a control sample.

\placefigure{fig:spectargetsonCMD}

The spectra were obtained with the Hectochelle multi-fiber, high-dispersion spectrograph \citep{Szentgyorgyi.98} on the MMT. Hectochelle can provide spectra for up to 240 sources in a single exposure over a limited filter-selected wavelength range. We reserved 55 fibers for observations of the sky and 18 fibers were unassigned. The observations were conducted over six separate nights as summarized in table~\ref{tab:specsummary}. The data from 03/03/2007 and 03/04/2007 have very high levels of sky contamination, so we were unable to extract usable spectra from them.

\placetable{tab:specsummary}

We used the RV31 filter which covers the range $5141.5-5310.5\AA$ at a resolution of 32000. The camera was read-out using $2 \times 3$ binning to improve the signal-to-noise of the faintest targets ($r \sim 21$). We used an exposure time of $20$ minutes. 

\section{Photometric Data Reduction}

The data reduction pipeline for photometric observations consists of five steps: 

\begin{enumerate}
\item CCD calibration including astrometry
\item PSF fitting photometry
\item Photometric calibration
\item Image subtraction time series photometry
\item Post-processing of light curves
\end{enumerate}

In this paper we will discuss the first three of these steps. The fourth step will be discussed in a separate paper on variable stars and the final step, which consists of removing real stellar variability as well as instrumental variations from the light curves for the purpose of searching for transits, will be discussed in a separate paper on the results of the transit survey.

\subsection{CCD Calibration}

The CCD calibration consists of subtracting the bias from each image extension using an overscan correction, trimming the overscan region from the image, applying a flat-field to the mosaic image, and merging the output from the two amplifiers on a given chip into a single image. The supplementary data (images used for constructing the CMDs and measuring the membership contamination) were calibrated using the standard routines in the IRAF MSCRED package\footnote{IRAF is distributed by the National Optical Astronomy Observatories, which is operated by the Association of Universities for Research in Astronomy, Inc., under agreement with the National Science Foundation.}. To reduce I/O overhead we performed the calibrations on the time series observations using our own implementation of the routines written in C.

To flatten our images we used observations of the twilight sky. Because conditions were acceptable for flat-fielding on only a handful of evenings, we constructed a single master flat-field in each filter from all available twilight images. 

Temporal variations in the flat-field over the course of the run may be responsible for some systematic variations in the light curves that must be removed before searching for transiting planets and other low amplitude phenomena. We did notice slight temporal variations in the gain between the two amplifiers that read out a given chip. To correct for this in calibrating the time series data we first applied the flat field and then scaled one of the amplifiers on each chip by the mode of the quotient of pixels to the left and right of the amplifier boundary. In performing the correction we did not match the chip to chip variations in the gain since the time series photometry is done independently for each chip.

To obtain the astrometric solution for the mosaic images we used the \emph{megawcs} program which is part of the Megacam reduction package\footnote{The Megacam reduction package is available from http://www.cfa.harvard.edu/\~{}bmcleod/Megared/}. We used the Two Micron All Sky Survey (2MASS) Point Source Catalog \citep{Skrutskie.06} as our astrometric reference.

\subsection{PSF Photometry}

To construct the CMDs for the cluster and the field off the cluster we performed PSF fitting photometry on a number of the images. We used the {\scshape Daophot} 2 and {\scshape Allstar} programs \citep{Stetson.87,Stetson.92} to do this. The process was run in batch mode because there were too many images to process by hand.

We followed a fairly standard procedure. We used {\scshape Daophot} to identify stars, choose candidate PSF stars, determine the PSF, and {\scshape Allstar} to fit the profile to all identified stars on the image. We then used the {\scshape Daogrow} program \citep{Stetson.90} to determine the aperture correction on the PSF stars. We did not search for additional stars on the residual images as the field is not terribly crowded and doing this typically yielded more spurious detections of artifacts in the residuals of bright stars than detections of additional stars. 

The data processed with PSF fitting photometry include all $g^{\prime}$ and $i^{\prime}$ cluster images, 56 $r^{\prime}$ mosaic images of the cluster taken during photometric conditions, images taken of the field off the cluster and the images taken of the Sloan fields.

In order to detect faint stars we also constructed a number of stacked images that we performed PSF fitting photometry on. Among these are the reference $r^{\prime}$ images of the primary field; the construction of these is done as part of the image subtraction photometry routine which will be described elsewhere. Briefly, the process consists of matching the background and seeing of $\sim 100$ of the best seeing images for each chip using the method of \citet{Alard.98} and \citet{Alard.00} before mean-combining them. Other stacked images include a set of short exposure $r^{\prime}$ images of the primary field taken through several magnitudes of extinction, sets of short exposure $g^{\prime}$ and $i^{\prime}$ images of the primary field, sets of long exposure $g^{\prime}$ and $i^{\prime}$ images of the primary field taken in good seeing conditions and sets of images of the field off the cluster. We stacked the short exposure images to provide overlap with the deeper images and thus tie the photometry of the brighter stars to that of the fainter stars. We used the high extinction $r^{\prime}$ images to obtain photometry for the bright stars because the short exposure $r^{\prime}$ images that we obtained did not provide adequate magnitude coverage. We note, therefore, that the $r^{\prime}$ photometry for the brightest stars is of poorer quality than the $g^{\prime}$ and $i^{\prime}$ photometry for these stars.

\subsection{Photometric Calibration}

To cover the main sequence from the turn-off at $r \sim 11$ down to M dwarf stars at $r \sim 23$ we tie together the photometry of the bright stars from the stacked short exposure images and the photometry of the faint stars from the stacked long exposure images. This instrumental photometry is then transformed to the Sloan 2.5 m system. Since the shallow and deep stacked images are composed of images obtained on a variety of non-photometric nights, the photometry from these images has to be matched to photometry of the cluster obtained during photometric conditions before transforming to the Sloan 2.5 m system. The steps involved may be summarized as follows:
\begin{enumerate}
\item Find the transformation from the instrumental PSF fitting magnitude measurements for a given photometric night to the Sloan 2.5 m system. The transformation includes zero-point, airmass and color terms.
\item Calculate the average instrumental magnitudes for target stars observed on that photometric night after correcting for differences between the exposure times and airmasses using the zero-point and airmass terms found in step 1.
\item Match the stacked short and long exposure images from non-photometric nights to the list from step 2 to get a full star list.
\item Apply the color terms in the transformation found in step 1 to the full list.
\end{enumerate}

\subsubsection{Photometric Calibration Step 1.}

Instrumental PSF fitting magnitude measurements are transformed to the Sloan 2.5 m system using observations taken of the equatorial Sloan field centered at 03:20:00, 00:00:00 (J2000). Conditions during the observing run were rarely photometric, as a result we were only able to constrain the airmass terms in the transformation from the observations of this field taken on 12/27/2005. We also attempted using the second Sloan field for calibration but found that the observations did not span a large enough range in airmass to constrain the airmass terms. After matching our star list for the images of the first Sloan field to the star list extracted from the SDSS Data Release 5, we fit a transformation of the form:
\begin{eqnarray}
g^{\prime} &=& g + a_{g}X + b_{g}(g - r) + z_{g,j} \nonumber \\
r^{\prime} &=& r + a_{r}X + b_{r}(r - i) + z_{r,j} \label{eqn:transformation}
 \\
i^{\prime} &=& i + a_{i}X + b_{i}(r - i) + z_{i,j} \nonumber
\end{eqnarray}
where $g^{\prime}$, $r^{\prime}$ and $i^{\prime}$ are instrumental magnitudes, $g$, $r$ and $i$ are Sloan magnitudes, $X$ is the airmass of the observation, $a_g$, $a_r$ and $a_i$ are the airmass terms to fit for, $b_g$, $b_r$ and $b_i$ are color terms to fit for and that we take to be the same for all chips in the mosaic and $z_{g,j}$, $z_{r,j}$ and $z_{i,j}$ are the zero-point terms for Chip $j$ that we fit for. We used a single color term for the entire mosaic because there were not enough stars in this field to constrain the color term for each chip independently. We used a single color term in each transformation since we found that including two color terms resulted in more scatter in the CMDs. The fit is performed only on stars with formal magnitude errors less than $0.02$ mag, in doing the fit we iteratively reject $3\sigma$ outliers. 

\begin{figure}[p]
\plotone{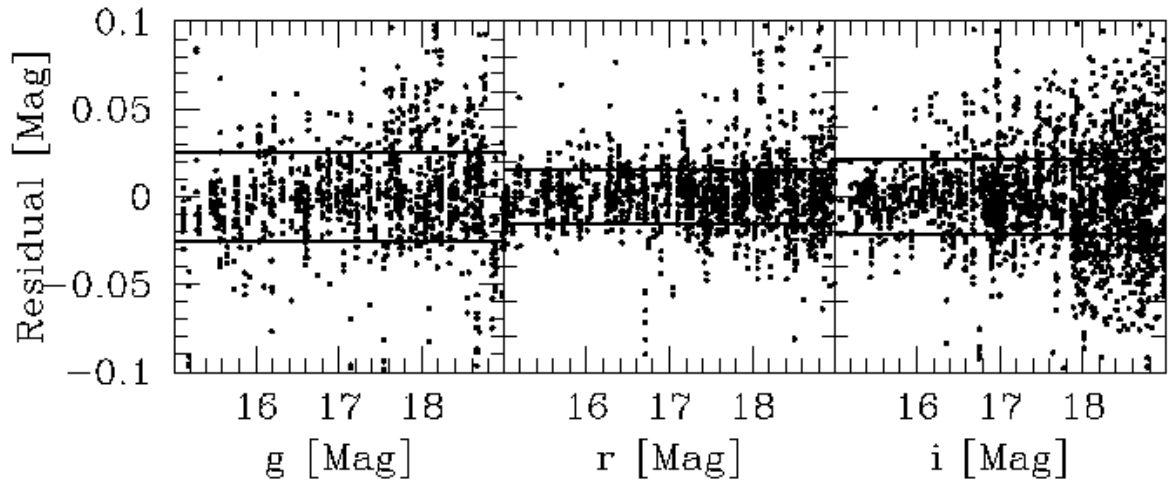}
\caption{Residuals after applying the transformation in equation~\ref{eqn:transformation} to the observed Sloan field. The left plot shows the $g$ data, the middle shows $r$ and the right shows $i$. The magnitudes on the x-axis are on the Sloan system. The lines show the RMS of the residuals: 0.025 mag in $g$, 0.016 mag in $r$ and 0.021 mag in $i$.}
\label{fig:photcalresid}
\end{figure}

\begin{figure}[p]
\plotone{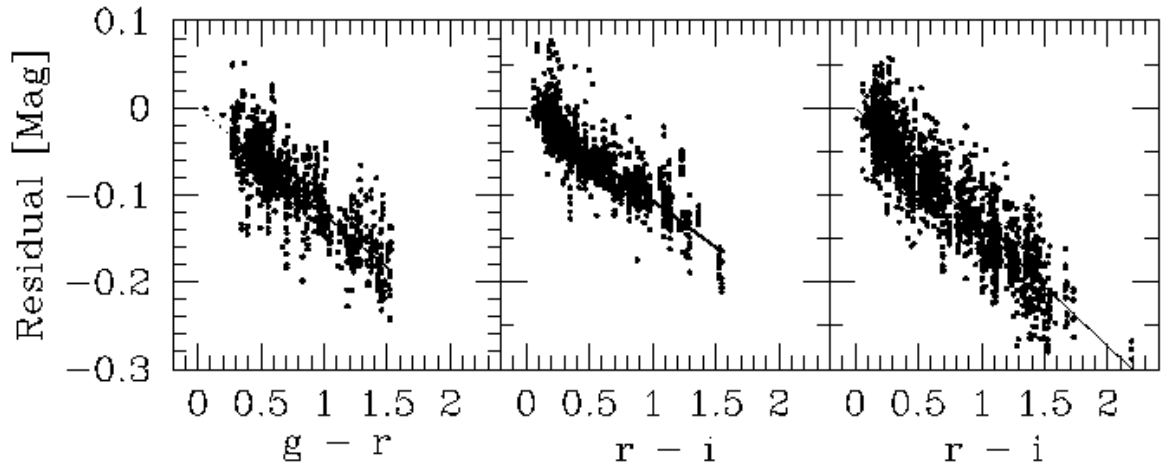}
\caption{Residuals after applying the transformation in equation~\ref{eqn:transformation} without the color term to the observed Sloan field. The left plot shows the $g$ data, the middle shows $r$ and the right shows $i$. The lines show the best fit color dependence for the transformation. For $g$ the slope is $-0.122$, for $r$ the slope is $-0.107$ and for $i$ the slope is $-0.137$.}
\label{fig:photcalcolor}
\end{figure}

The parameters that we determined, together with their standard errors, are listed in tables~\ref{tab:photcalparams} and~\ref{tab:photcalparams2}; the listed zero points are adjusted to have a mean value of 0 over all the chips. The RMS of the residuals after applying the transformation in equation~\ref{eqn:transformation} to the Sloan data are 0.025 mag in $g^{\prime}$, 0.016 mag in $r^{\prime}$ and 0.021 mag in $i^{\prime}$. We take these as estimates of the systematic error in our transformation to the Sloan system. In figure~\ref{fig:photcalresid} we plot the residuals after applying the transformation to the Sloan data. In figure~\ref{fig:photcalcolor} we show the color dependence of the transformation by plotting the residuals after applying everything except the color terms in the transformation.

\placetable{tab:photcalparams}

\placefigure{fig:photcalresid}

\placefigure{fig:photcalcolor}

\subsubsection{Photometric Calibration Step 2.}

After fitting for the parameters, we invert the transformation in equation~\ref{eqn:transformation} to yield $g$, $r$ and $i$ for a star as a function of $g^{\prime}$, $r^{\prime}$, $i^{\prime}$, $X$ and the chip number. To apply this transformation to the M37 data we first define $\tilde{g} = g^{\prime} - a_{g}X - z_{g,j}$ and similarly for $\tilde{r}$ and $\tilde{i}$. These corrections are applied to all magnitude measurements of stars in the M37 field made during photometric conditions on 12/27/2005. We then match sources from different images using a radius of $0\farcs05$ and calculate the average $\tilde{g}$, $\tilde{r}$ and $\tilde{i}$ values for each source. We use the resulting star list as a template to tie together the higher precision, deeper photometry from the stacked images.

\subsubsection{Photometric Calibration Step 3.}

We match the star lists from each of the stacked images to the template list and calculate the magnitude offset between the matched stars for each chip. This offset is then applied to the stacked image magnitudes. For the bright stacked $g^{\prime}$ observations the scatter on this transformation is typically less than $0.02$ mag, for the deeper stacked $g^{\prime}$ observations the scatter is typically less than $0.01$ mag. The corresponding values for $r^{\prime}$ are $0.02$ mag and $0.008$ mag, while for $i^{\prime}$ the values are $0.025$ mag and $0.009$ mag. The stacked image star lists are then combined to form a full list of stars with $\tilde{g}$, $\tilde{r}$ and $\tilde{i}$ magnitudes. In combining the stacked star lists, when a source is detected in both the shallow and deep images we take the error weighted average of the two magnitudes if the difference between them is less than $0.05$ mag and the instrumental magnitude measured on the deeper image is less than $12$, otherwise we ignore the magnitude from the shallow image. 

\subsubsection{Photometric Calibration Step 4}

The color transformations are applied to the full star list to yield a final list with magnitudes on the Sloan system. Only sources detected in all three filters are transformed. This final star list has 16483 sources with $10.59 < g < 26.76$, $9.96 < r < 24.71$ and $9.43 < i < 24.30$. Table~\ref{tab:phot_catalog} provides the first few lines of our photometric catalog, the full version is available in the electronic edition of the journal - note that we have kept our own numbering system in this table.

\subsubsection{Calibrating the Off-Cluster Images}

The observations of the field off the cluster were not taken on the same nights as the observations of the Sloan fields, though they were taken under photometric conditions. To convert these observations to the Sloan system we determine the transformation from instrumental magnitudes on the night in question to Sloan magnitudes using observations of M37. Since we did not have enough observations to constrain the airmass term in the transformation we use the airmass term that was determined on 12/27/2005; we expect that the error from making this assumption should be less than $\sim 0.02$ mag.

\section{Spectroscopic Results}

\subsection{Spectroscopic Data Reduction}

The Telescope Data Center (TDC) at the Smithsonian Astrophysical Observatory (SAO) provides reduced Hectochelle spectra to observers. The reduction pipeline is described on the TDC website\footnote{http://tdc-www.harvard.edu/instruments/hectochelle/pipeline.html}. For sources fainter than $r \sim 17$ we found that sky contamination was significant. To correct for this we fit, at each wavelength, a third order polynomial in the $x$ and $y$ fiber positions to the 55 sky spectra. This provides an estimate for the sky spectrum at the location of each target fiber which we then subtract. We use the reduced spectra to classify the stars and measure their radial velocities as described in the following subsections.

\subsection{Spectroscopic Determination of $T_{eff}$, $[Fe/H]$ and $v\sin i$}

Despite the fact that our Hectochelle observations cover only a $169\AA$ range, there are a number of absorption lines in this range that can be used to determine spectroscopic parameters ($T_{eff}$, $[Fe/H]$, $\log(g)$ and $v\sin i$) for the observed stars. The observed wavelength range includes the MgI triplet lines as well as many weaker lines of neutral Mg, Fe, Ti, Cr, and Ni. The temperature is the variable that most strongly controls the line strength. The sensitivity arises from the exponential and power dependencies with temperature in the excitation-ionization processes. The MgI lines are also very sensitive to gravity with their wings dominated by Stark and van der Waals broadening. The dependence on gravity of the weaker lines depends on their ionization state. Metallicity affects all lines.

We classify the stars following the procedure due to Meibom et al. (2007, in preparation) that has been developed specifically for the Hectochelle RV31 filter. The procedure is to cross-correlate the spectra against templates drawn from a grid of ATLAS 9 model atmospheres and computed using the companion program SYNTHE \citep{Kurucz.93}. The current library has a total of 51,359 spectra over effective temperatures ($T_{eff}$) from $3500~{\rm K}-9750~{\rm K}$ in steps of $250~{\rm K}$, over gravity from $\log(g)$ of $0.0-5.0$ in steps of $0.5$, over metallicity ($[Fe/H]$) from -2.5 to +0.5 in steps of 0.5, and over rotation velocity from $0-200$ km/s. At this time $[\alpha/Fe]$ is set to 0.0, microturbulence to 2 km/s and macroturbulence to 1 km/s. 

The cross-correlation is performed using the \emph{xcsao} routine in the {\scshape Iraf} \emph{rvsao} package \citep{Kurtz.98}. For each spectrum we determine the $T_{eff}$, $\log(g)$, $v\sin i$ and $[M/H]$ which yields the highest cross-correlation R value. The peak values are interpolated within the grid. Note that we used model atmospheres without $\alpha$-enhancement, so $[Fe/H] = [M/H]$. We have also attempted fixing $T_{eff}$ while varying the other three parameters as well as fixing $\log(g)$ to $4.5$ (assuming all stars are dwarfs) while varying the other parameters. To fix $T_{eff}$ for each star we use the star's $B-V$ color from \citet{Kalirai.01}, a reddening of $E(B-V) = 0.227$ and the theoretical color-effective temperature relation from the $[M/H] = 0.045$, 485 Myr YREC isochrone (see \S 5.2). 

We find that the spectra are not of high enough quality to independently determine all four parameters; when we attempt to do so the resulting $T_{eff}-\log(g)$ and $(B-V)-T_{eff}$ relations are inconsistent with theoretical expectations. Similarly, when using $(B-V)$ to fix $T_{eff}$ the resulting $T_{eff}-\log(g)$ relation is inconsistent with the theoretical relation. We find, however, that when we fix $\log(g) = 4.5$ the $(B-V)-T_{eff}$ relation is in good agreement with the theoretical relation. We therefore choose to fix $\log(g) = 4.5$, which is a good assumption for the cluster members as well as for the field stars which are predominately background dwarfs. 

\begin{figure}[p]
\plotone{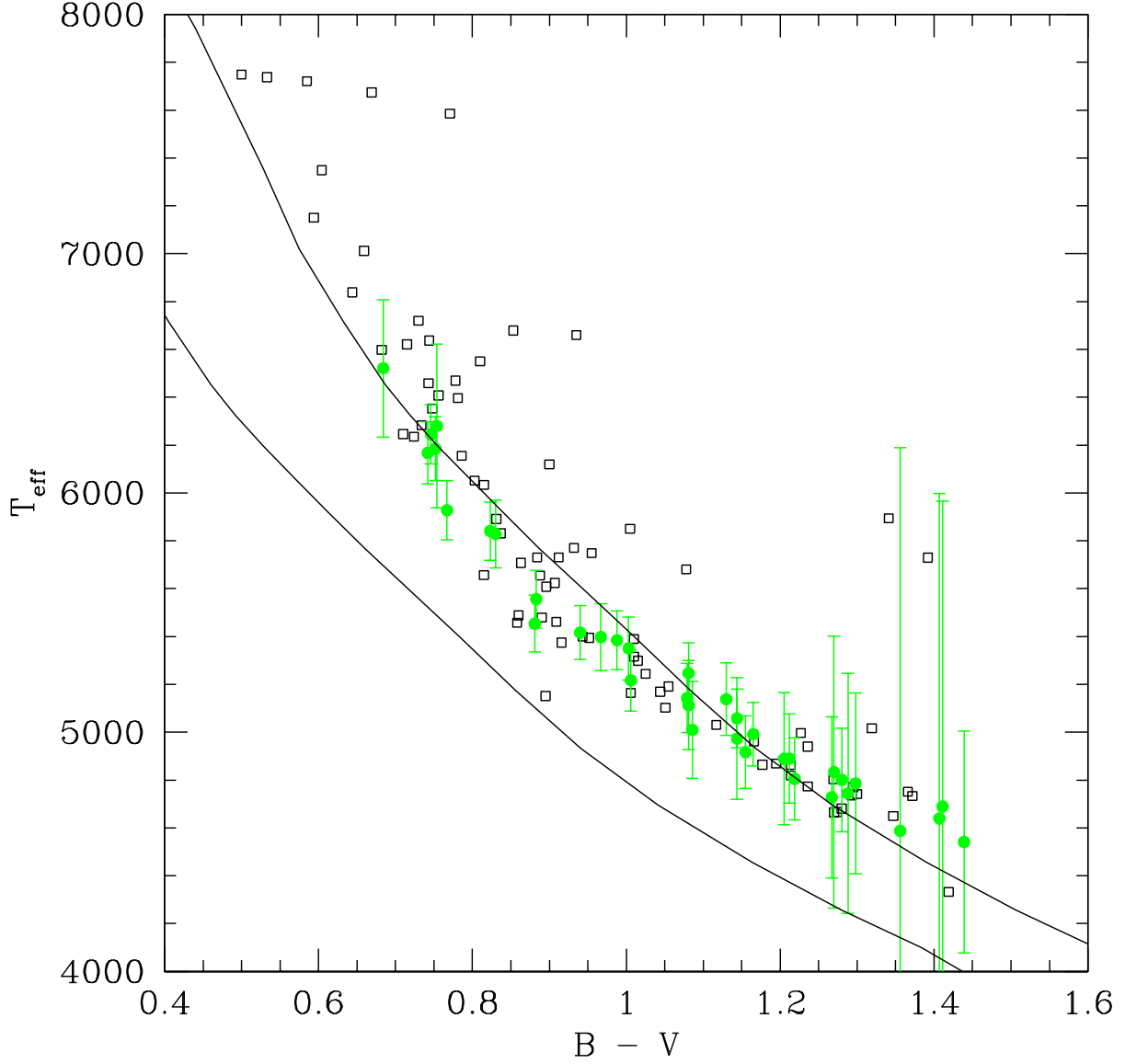}
\caption{The $B-V$ color versus spectroscopically inferred temperature for candidate cluster members that pass both photometric and radial velocity selections (filled points) and other stars (open points). In this case $T_{eff}$, $v\sin i$ and $[M/H]$ have been allowed to vary in the cross-correlation while $\log(g)$ is fixed to 4.5. The lines show the theoretical color-temperature relations for $[M/H] = 0.045$ from the YREC isochrones. The left line is for zero reddening, while the right line is for $E(B-V) = 0.227$. Note that as expected, non-members tend to show greater reddening than members.}
\label{fig:BVTeff_fixlogg}
\end{figure}

\begin{figure}[p]
\plotone{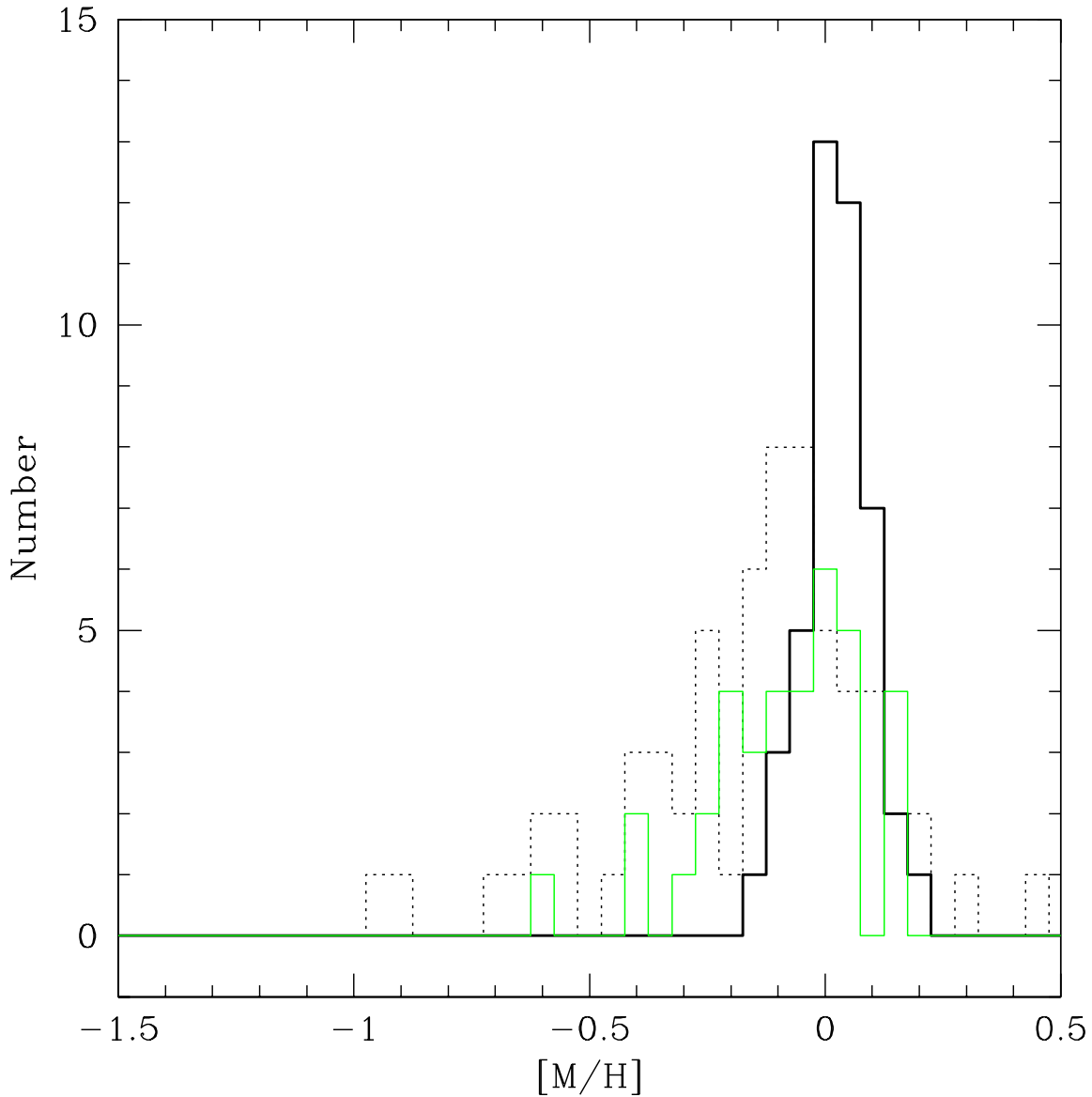}
\caption{Histogram of metallicities for candidate cluster members that pass both photometric and radial velocity selections (dark solid line), stars that pass the photometric membership selection but fail the radial velocity selection (light solid line) and other stars (dot-dashed line). In this case $T_{eff}$, $v\sin i$ and $[M/H]$ have all been allowed to vary in the cross-correlation while $\log(g)$ is fixed to 4.5. Cluster members show a peak at around solar metallicity while non-members show a broader distribution with a peak toward lower metallicity.}
\label{fig:HistMet_fixlogg}
\end{figure}

Figure~\ref{fig:BVTeff_fixlogg} shows the $B-V$ vs. $T_{eff}$ relation when fixing $\log(g) = 4.5$. The cluster sequence is in good agreement with the expected value assuming $E(B-V) = 0.227$. Figure~\ref{fig:HistMet_fixlogg} shows the resulting metallicity histogram for photometrically selected cluster members (\S 5.1) that have a radial velocity within 3$\sigma$ of the cluster systemic velocity (see below), photometrically selected cluster members that have a radial velocity that is discrepant from the systemic velocity by more than $3\sigma$ and non-members. The cluster members that pass the radial velocity cut show a distribution that is strongly peaked at $[M/H] \sim 0.0$ while the other stars show a broader distribution that is peaked at a somewhat lower metallicity. The weighted mean of the cluster metallicity is $[M/H] = 0.02 \pm 0.04$ (excluding stars with $T_{eff} < 4500$ K or $\sigma_{[M/H]} > 0.5$), which is consistent with the spectroscopic determination of $[Fe/H] = 0.05 \pm 0.05$ by \citet{Marshall.05}.

\placefigure{fig:BVTeff_fixlogg}

\placefigure{fig:HistMet_fixlogg}

Since the measured values of $T_{eff}$, $[M/H]$ and $v\sin i$ are correlated with $\log(g)$, it is important to estimate the systematic error on these parameters that results from fixing $\log(g) = 4.5$. From the theoretical YREC isochrones, we expect that the observed cluster members should have $\log(g)$ ranging from $4.3$ at the bright end to $4.7$ at the faint end. By performing the cross-correlation with $\log(g) = 4.0$ and $\log(g) = 5.0$ we find that fixing $\log(g) = 4.5$ results in a systematic error per point in $T_{eff}$ of $\pm 100~{\rm K}$, a systematic error in $[M/H]$ of $\pm 0.075~{\rm dex}$ and negligible systematic errors in $v\sin i$. Note that since the observed stars are expected to have $\log(g)$ evenly distributed about $\log(g) = 4.5$ we do not expect these systematic errors to significantly affect the estimates of the cluster parameters. To determine the total error for each star we add the systematic errors in quadrature to the errors estimated by taking the standard deviation of the parameters measured independently on each of the four nights.

\subsection{Cluster Radial Velocity}

\begin{figure}[p]
\plotone{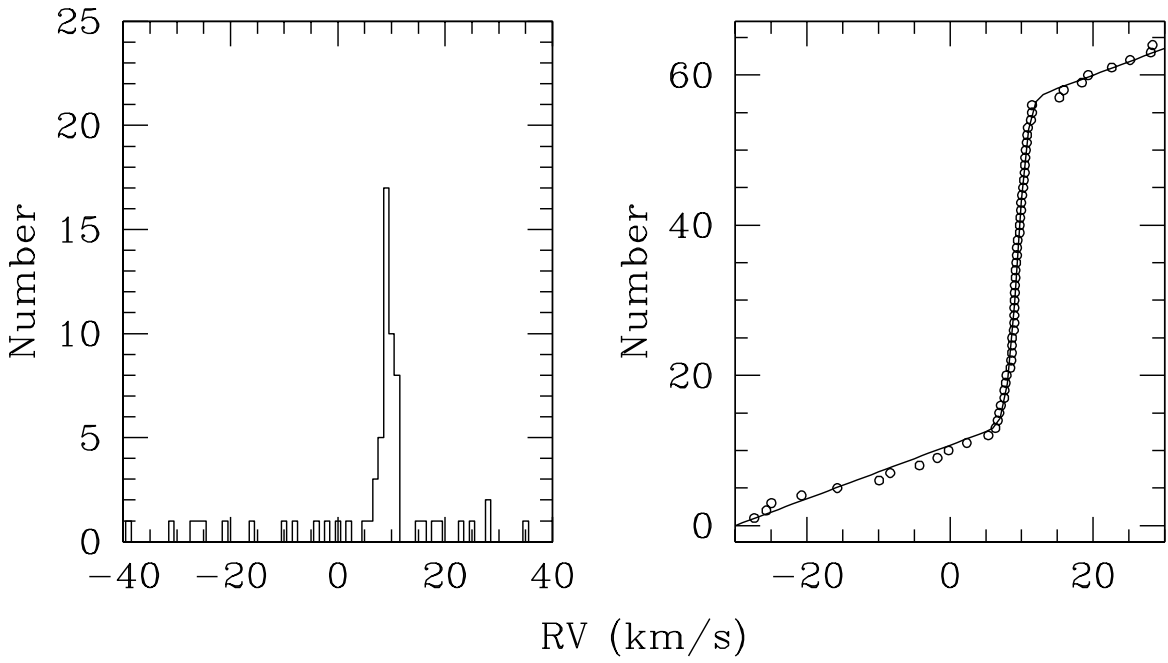}
\caption{Left: Histogram of radial velocities for stars near the $gri$ main sequences of M37. Right: The observed cumulative distribution of radial velocities (open points) is plotted together with the best fit model (equation~\ref{eqn:RVcumdist}; solid line).}
\label{fig:RVhistplot}
\end{figure}

The radial velocity for each star is computed on each night using \emph{xcsao} with the best average synthetic template for the star determined in the previous subsection. Figure~\ref{fig:RVhistplot} shows the histogram of average radial velocities for 74 bright target stars near the main sequence in $gri$ (as per \S 5.1). The radial velocities have been corrected to the barycenter of the solar system. The distribution shows a clear peak around $\sim 9$ km/s which we take to be the systemic RV of the cluster. 

To determine the systemic radial velocity and velocity dispersion of the cluster we calculate the unnormalized cumulative radial velocity distribution function (CDF; the value of the CDF at point $x$ is the number of measurements with $RV$ less than or equal to $x$) and fit to it a function of the form:
\begin{equation}
\Phi(RV) = N_{f}(RV-RV_{min}) + N_{c}({\rm erf}(\frac{RV - RV_{0}}{\sqrt{2}\sigma_{RV}}) - {\rm erf}(\frac{RV_{min} - RV_{0}}{\sqrt{2}\sigma_{RV}}))
\label{eqn:RVcumdist}
\end{equation}
where the first term is a uniform distribution to model the field stars and the second term is a Gaussian distribution to model the cluster. Here $RV_{0}$ is the systemic velocity of the cluster, $\sigma_{RV}$ is the projected velocity dispersion of the cluster, $RV_{min}$ is the minimum $RV$ in the fitting range and $N_{f}$ and $N_{c}$ are normalization parameters. The fit is performed over the interval $-30~{\rm km/s} < RV < 30~{\rm km/s}$ where the field stars appear to roughly follow a uniform distribution. The best fit is plotted in figure~\ref{fig:RVhistplot}. We find $RV_{0} = 9.4 \pm 0.2~{\rm km/s}$ and $\sigma_{RV} = 1.2 \pm 0.2~{\rm km/s}$ where the errors are set equal to the standard deviation of the measured parameters from 1000 bootstrap iterations. Note that if the underlying radial velocity distribution for the cluster is a delta function, the expected RMS of the observed distribution given the uncertainties on the RV measurements (taken to be the standard deviation of the RV values for each star) is $0.5~{\rm km/s}$. Though the measured dispersion is most likely dominated by the true internal velocity dispersion of the cluster, it may have a contribution from binaries which are not filtered from the small set of observations.

\placefigure{fig:RVhistplot}

For comparison, \citet{Mermilliod.96} find the mean velocity of $7.68 \pm 0.17$ km/s and a dispersion of $0.92$ km/s from 30 non-variable red giants. Part of the discrepancy in the systemic RV measurements is due to the different gravitational redshifts between main sequence and giant stars \citep[e.g.][]{Gonzalez.01}. The gravitational redshift of a star is $0.636 (M/M_{odot})/(R/R_{\odot})$ km/s, so that a $1 M_{\odot}$ star in M37 will have a gravitational redshift of $\sim 0.7$ km/s, whereas from the Padova isochrones (see \S 5.2.4) we expect the red clump stars in the cluster to have a redshift of $\sim 0.1$ km/s. Additionally, convective motions in stellar atmospheres can result in a net blueshift of several hundred m/s in spectral lines \citep{Dravins.81} with some indications that the effect is stronger in giants than dwarfs \citep{AllendePrieto.02}. Both of these effects yield a lower systemic RV for giants than for dwarfs. 

If the cluster is in energy equipartition, then one would expect the velocity dispersion to decrease with increasing stellar mass such that $\sigma \propto M^{-0.5}$. Assuming the giants have masses of $\sim 2.5 M_{\odot}$ whereas the main sequence stars we observed have masses of $\sim 1.0 M_{\odot}$, then one would expect a velocity dispersion ratio of $\sigma_{dwarf}/\sigma_{giant} \sim 1.6$ which is slightly greater than the observed ratio of $1.3 \pm 0.2$.  

Integrating the contribution to equation~\ref{eqn:RVcumdist} from the cluster, we expect that 42 of the 74 stars in the sample are cluster members, this yields a contamination frequency of $43\%$. If we also require the stars to be near the $B-V$ main sequence this frequency falls to $25\%$.

\section{Cluster Parameters}

The color magnitude diagrams together with the spectra can be used to refine measurements of the cluster metallicity, distance, reddening and age. To do this we first must select a sample of probable cluster members. 

\subsection{Membership Selection}

Probable \emph{single star} cluster members to use in determining the fundamental cluster parameters are selected photometrically. To do the selection we use $B$ and $V$ photometry from \citet{Kalirai.01}, the $g$, $r$ and $i$ photometry presented in this paper, and the $I_{C}$ photometry transformed from $r$ and $i$ (see below). The selection is done following a procedure similar to \citet[][Hereafter A07]{An.07}. We identify, by eye, a fiducial main sequence and red clump in each of the $B-V$ vs. $V$, $g-r$ vs. $r$, $g-i$ vs. $i$ and $V-I_{C}$ vs. $V$ CMDs. We used $V-I_{C}$ rather than $V-g$, $V-r$ or $V-i$ because $V-I_{C}$ is the CMD used for fitting (See \S 5.2.1). For each star $i$, in each CMD, we determine the magnitude pairs $M_{1}$ and $M_{2}$, interpolated within the fiducial sequence, that minimizes
\begin{equation}
\chi_{i}^{2} = \frac{(m_{1,i} - M_{1})^{2}}{\sigma_{m,1,i}^2} + \frac{(m_{2,i} - M_{2})^{2}}{\sigma_{m,2,i}^2}.
\label{eqn:chi2star}
\end{equation}
Here $m_{1,i}$ and $m_{2,i}$ are the measured magnitudes for star $i$ in the two filters and $\sigma_{m,1,i}$ and $\sigma_{m,2,i}$ are the corresponding errors. We then select stars with $\chi_{i}^{2} < 150$ in all four CMDs, and $\sigma_{m,i} < 0.03$ mag in every filter. The selection cutoff is fairly high as the photometric errors are likely underestimated. The cutoff was determined by eye, using a smaller value leaves an over-density of stars on either edge of the selected sequence while using a larger value begins picking up what we consider to be the obvious field star population. We did experiment with using an iterative procedure to automatically define the fiducial main sequence as described in A07, but we found that this did not seem to be very robust for the high field density of this cluster. 

\begin{figure}[p]
\plotone{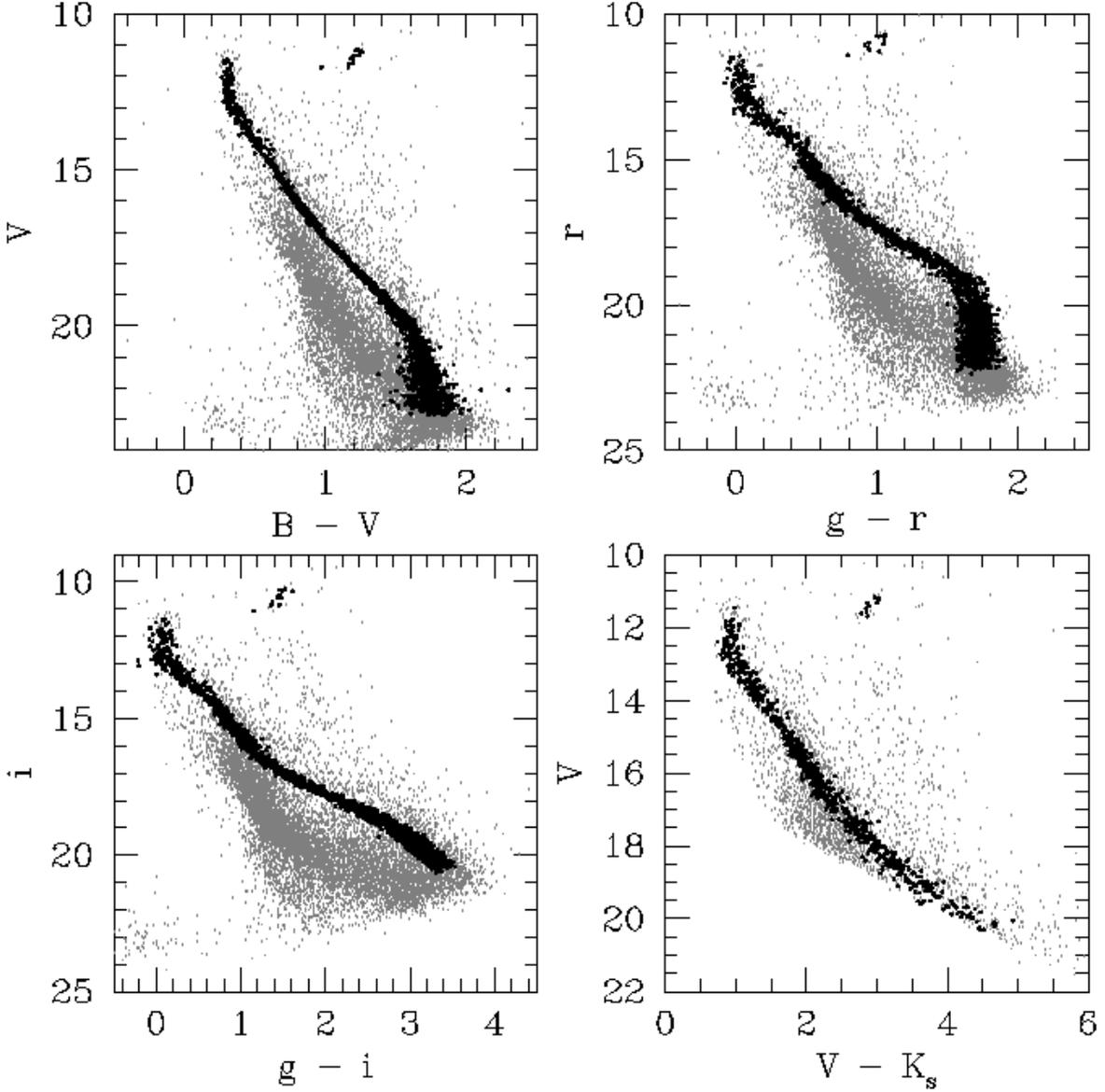}
\caption{The photometrically selected candidate cluster members (dark points) are shown on $B-V$, $g-r$, $g-i$ and $V-K_{S}$ CMDs. The light points show all other stars. We use subsets of these stars to determine the cluster parameters (\S 5.2.2). Note that the faint magnitude cut on the candidate members is due to a cut on the photometric precision made in selecting these stars. The $V-K_{S}$ CMD is not used to select candidates, the fact that none of the selected stars are outliers in this diagram shows that the selection is robust.}
\label{fig:selectMS}
\end{figure}

The resulting selections are shown in figure~\ref{fig:selectMS}. Note that in this figure we also plot the $V-K_{S}$ vs. $V$ CMD, with $K_{S}$ from 2MASS to show that the selection is robust; to save space the $V-I_{C}$ vs. $V$ CMD is omitted. 

A total of 1473 stars are selected by this method. Of the 53 stars with spectroscopy that have $RV$ within $3\sigma$ of the cluster's systemic $RV$, 36 (68\%) are selected by this method. One of the unselected stars does not have $BV$ photometry, five lie at the edge of the selected main sequence and would be included if the photometric selection criteria were slightly relaxed, one has a $B$ magnitude that is most likely in error ($B-V > 3.0$) and ten lie well away from the main sequence. 

\placefigure{fig:selectMS}

\subsection{Isochrone Fitting}

Main sequence fitting is one of the oldest, and most robust techniques for determining the distance and reddening to star clusters when these parameters cannot be measured directly (e.g.~via parallax). When the metallicity of a cluster is well known distances in parsecs accurate to better than a few percent can be obtained by this technique A07. As shown by \citet{Pinsonneault.04} theoretical isochrones generated with the Yale Rotating Evolutionary Code \citep{Sills.00} successfully reproduce the $L-T_{eff}-M$ relations for an eclipsing binary in the Hyades and are in agreement with the spectroscopically determined $T_{eff}$ measurements for Hyades stars with good parallax measurements; however, there appear to be systematic discrepancies between the theoretical and observed Hyades color-magnitude main sequence that are most likely due to errors in the theoretical color-$T_{eff}$ relations. Using an empirical Hyades-based correction to the \citet{Lejeune.97,Lejeune.98} color-$T_{eff}$, A07 have constructed a set of main sequence isochrones that successfully reproduce the CMDs for several nearby open clusters. Moreover, they show that metallicities better than $0.1$ dex and color excesses accurate to a few hundredths of a magnitude, including systematic errors, may be obtained from $BVI_{C}K_{S}$ photometry alone using these isochrones.

As mentioned in the introduction, there have been a number of previous determinations of the fundamental cluster parameters via isochrone fitting to CMDs. Table~\ref{tab:sumprevisocresults} gives a summary of the previous results. 

In most of these determinations the metallicity has been assumed to either be solar or has been taken from a different source. \citet{Mermilliod.96} attempted fitting a sub-solar metallicity isochrone as well as a solar metallicity isochrone. The parameters which they report are for the solar metallicity isochrone, though they note that a slightly sub-solar metallicity would provide a better fit to the red giants. \citet{Nilakshi.02} also attempted fitting both $Z = 0.008$ and $Z = 0.020$ isochrones by eye. They found that the $Z = 0.008$ isochrone provided a better fit. \citet{Sarajedini.04} adopted a metallicity value of $[Fe/H] = +0.09$ and reddening of $E(B-V) = 0.27 \pm 0.03$ from \citet{Twarog.97}. To determine the distance to the cluster they used an empirical color-magnitude-metallicity relation. \citet{Kalirai.05} adopted a sub-solar metallicity of $Z = 0.011 \pm 0.001$ and a fixed $E(B-V) = 0.23 \pm 0.01$ based on an unpublished spectroscopic determination by C.~Deliyannis. Note that the listed ages for \citet{Kalirai.01} and \citet{Kalirai.05} are based on isochrone fitting to the turnoff and clump stars, in both papers they have also determined comparable ages by measuring the end of the white dwarf cooling sequence. Also note that the distance and reddening determined by \citet{Kalirai.01} is based on fitting the fiducial Hyades main sequence to the M37 main sequence. Finally \citet{Kang.07} assumed solar metallicity.

\placetable{tab:sumprevisocresults}

\subsubsection{Isochrone Fitting - Transformation to $I_{C}$}

\begin{figure}[p]
\plotone{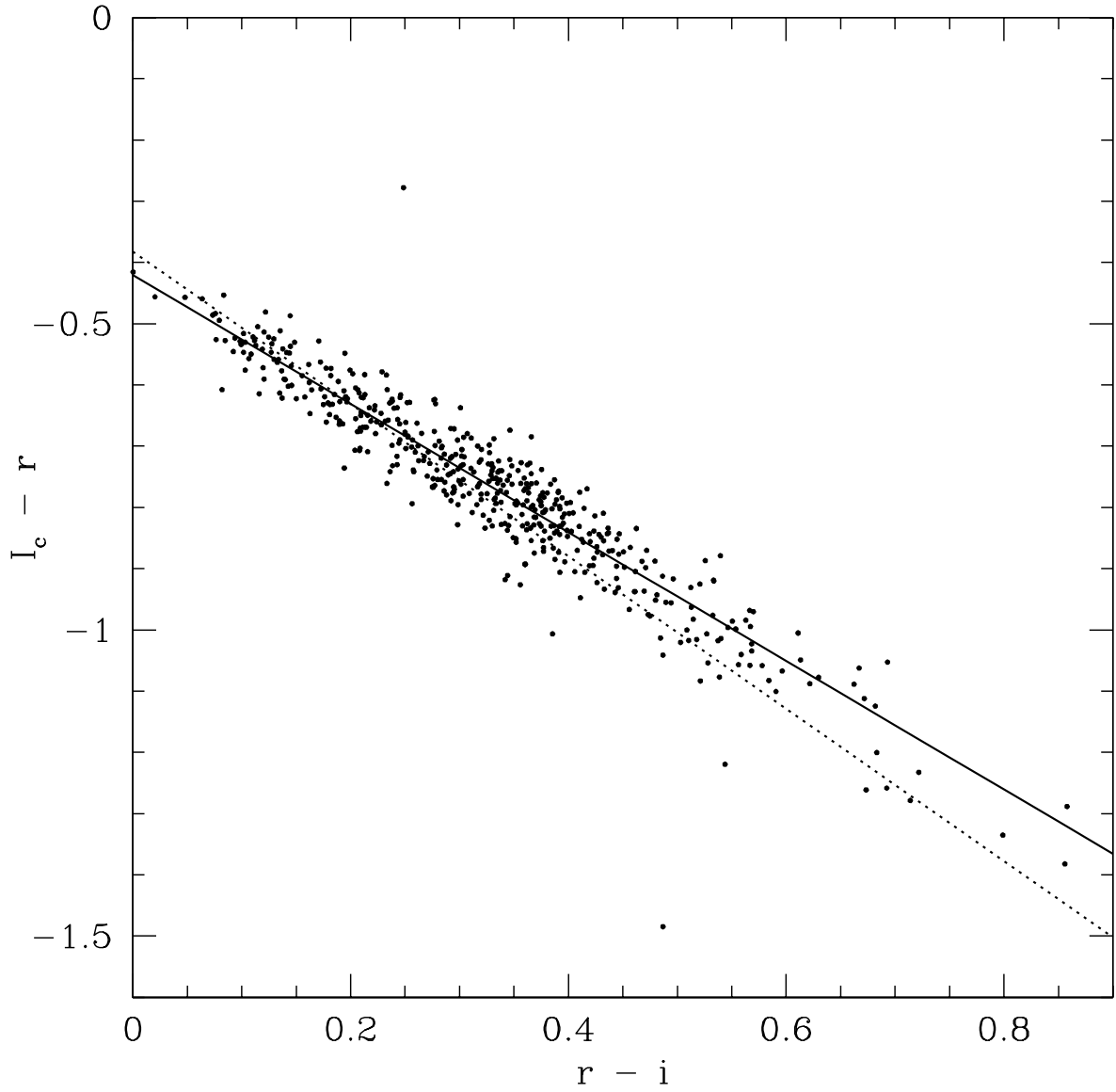}
\caption{Transformation between the $ri$ photometry presented in this paper and the $I_{C}$ photometry from \citet{Nilakshi.02}. The solid line shows our adopted transformation. The dashed line shows the Lupton (2005) transformation provided on the SDSS photometric equations website.}
\label{fig:ritoI}
\end{figure}

Because we have obtained deeper photometry in the red filters ($r$ and $i$) than has been previously available for this cluster, it is worthwhile to perform another independent main sequence fitting. To fit the main sequence we use the A07 empirically calibrated isochrone grid (we refer to these as the YREC isochrones in the remainder of the paper). This grid can be used to fit $B-V$, $V-I_{C}$ and $V-K_{S}$ CMDs. We take the $B$ and $V$ photometry from \citet{Kalirai.01} and the $K_{S}$ photometry from 2MASS. We transform our $r$ and $i$ photometry into $I_{C}$ using the $I_{C}$ photometry from \citet{Nilakshi.02}. We find
\begin{equation}
I_{C} = r - (1.050 \pm 0.099) \times (r-i) - (0.421 \pm 0.034)
\end{equation}
Figure~\ref{fig:ritoI} shows this transformation, together with the Lupton (2005) transformation listed on the SDSS photometric equations website\footnote{http://www.sdss.org/dr5/algorithms/sdssUBVRITransform.html}.

\placefigure{fig:ritoI}

\subsubsection{Isochrone Fitting - Photometric Errors}

Before performing the fit we first ensure that the photometric errors adequately describe the scatter in the data. There may be several reasons why the observed scatter in the main sequence could be larger than expected based on the photometric errors. Binarity and stellar variability both act to broaden the observed sequence, systematic errors between the chip to chip zero-point terms may also add scatter to the observed sequence, and finally the errors on the measured flux may be too optimistic (due to systematic errors in the model PSF for example). These errors effectively produce a constant error term which can be added in quadrature to the formal photometric errors. To account for this we find, for each object, the $V$, $B$, $I_{C}$ and $K_{S}$ values interpolated within the fiducial main sequences that minimize $\chi_{i}^{2}$ as in equation~\ref{eqn:chi2star}, where we now include all four magnitudes in the sum. We then find constant error terms $\sigma_{0,V}$, $\sigma_{0,B}$, $\sigma_{0,I_C}$ and $\sigma_{0,K_{S}}$ to add in quadrature to the listed errors such that 
\begin{equation}
\sum_{i=1}^{N} \chi_{m,i}^{2} = \frac{3}{4}N
\end{equation}
for each magnitude. Here $N$ is the number of stars, $\chi_{m,i}^{2}$ is the contribution to $\chi_{i}^{2}$ in equation~\ref{eqn:chi2star} from filter $m$, and the factor of 3/4 is needed to account for the parameter used for each star to fit its position within the fiducial main sequence. We find $\sigma_{0,V} = 0.008$, $\sigma_{0,B} = 0.017$, $\sigma_{0,I_{C}} = 0.047$ and $\sigma_{0,K_{S}} = 0.094$ magnitudes when limited to main sequence stars with $V > 12.25$, $B-V < 1.0$. 

\subsubsection{Isochrone Fitting - YREC Isochrones}

To fit the main sequence with the YREC isochrones we minimize the $\chi^{2}$ statistic:
\begin{equation}
\chi^{2} = \sum_{i=1}^{N} \chi_{i}^{2}
\end{equation}
where the $\chi_{i}^{2}$ values for each star is given by:
\begin{equation}
\chi_{i}^{2} = (\sum_{j=1}^{4}\frac{(M_{j} - m_{i,j})^{2}}{\sigma_{m,i,j}^{2}}) + \frac{(T_{eff} - T_{eff,i})^{2}}{\sigma_{T,i}^{2}} + \frac{([M/H] - [M/H]_{i})^{2}}{\sigma_{[M/H],i}^{2}}.
\end{equation}
In this equation terms with $i$ subscripts correspond to measured values, the sum on $j$ is over the filters $B$, $V$, $I_{C}$, and $K_{S}$, the theoretical apparent magnitudes $M_{j}$, and temperatures $T_{eff}$ are interpolated functions of the stellar mass, metallicity, and age. The theoretical apparent magnitudes are also functions of the apparent distance modulus $(m-M)_{V}$ and color excesses $E(B-V)$, $E(V-I_{C})$ and $E(V-K_{S})$. The spectroscopic temperature and metallicity measurements are only included for stars with those values. 

For a given set of cluster parameters (age, $[M/H]$, $(m-M)_{V}$, and color excesses) we choose the mass for each star that minimizes $\chi_{i}^{2}$. We then vary the cluster parameters using the downhill simplex method \citep{Nelder.65, Press.92} to minimize the total $\chi^{2}$. We restrict the fit to stars with $V > 12.25$, $B-V < 1.0$; the cut on the brighter stars is necessary because the YREC isochrones are not extended beyond the turnoff and including the stars at the turnoff would make the fit sensitive to extrapolation ambiguities, the cut on the red stars is necessary because theoretical isochrones are known to fail for K and M dwarfs \citep[e.g.][]{Baraffe.98}. Rather than imposing a reddening law to relate the color excesses, we allow them to vary independently, this allows for systematic errors in the photometric zero-points. 

\begin{figure}[p]
\plotone{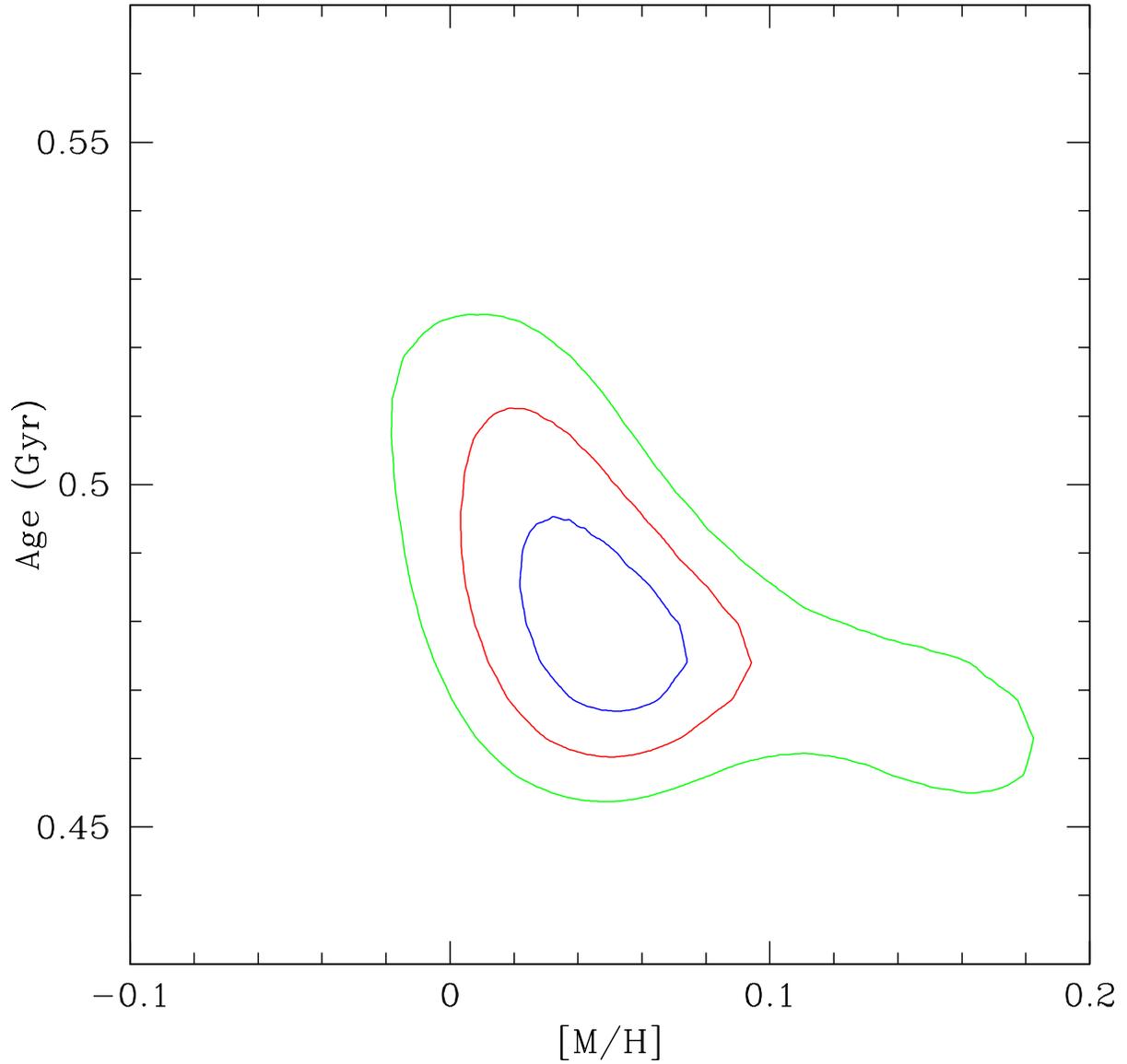}
\caption{Contours of constant $\Delta \chi^{2}$ in the $[M/H]$ vs. $t$ plane for a fit to the upper portion of the M37 main sequence allowing $(m-M)_{V}$, $E(B-V)$, $E(V-I_{C})$ and $E(V-K_{S})$ to vary independently. Contours are shown for $\Delta \chi^{2} = 2.30$, 6.17 and 11.8 which correspond to the $68.3\%$, $95.4\%$ and $99.7\%$ confidence levels.}
\label{fig:MSfitchi2_1}
\end{figure}

We first perform the fit without any spectroscopic constraints. The resulting parameters are: $t = 478 \pm 16$ Myr, $[M/H] = +0.049 \pm 0.031$, $(m-M)_{V} = 11.600 \pm 0.048$, $E(B-V) = 0.231 \pm 0.010$, $E(V-I_{C}) = 0.359 \pm 0.009$, and $E(V-K_{S}) = 0.705 \pm 0.021$. The errors are $1\sigma$ values from 1000 bootstrap iterations and do not account for possible errors in the model, systematic errors in the photometry or fitting methods, or biases that may result from neglecting effects such as binarity. Figure~\ref{fig:MSfitchi2_1} shows contours of constant $\Delta \chi^{2}$ projected onto the $[M/H]$ vs. $t$ plane. Contours are shown for $\Delta \chi^{2} = 2.30$, 6.17 and 11.8 which correspond to the $68.3\%$, $95.4\%$ and $99.7\%$ confidence levels for 2 degrees of freedom. We note that there is a higher metallicity secondary minimum between the 2 and $3\sigma$ levels which provides a better fit to the lower main sequence but a poorer fit to the upper main sequence.

\placefigure{fig:MSfitchi2_1}

\begin{figure}[p]
\plotone{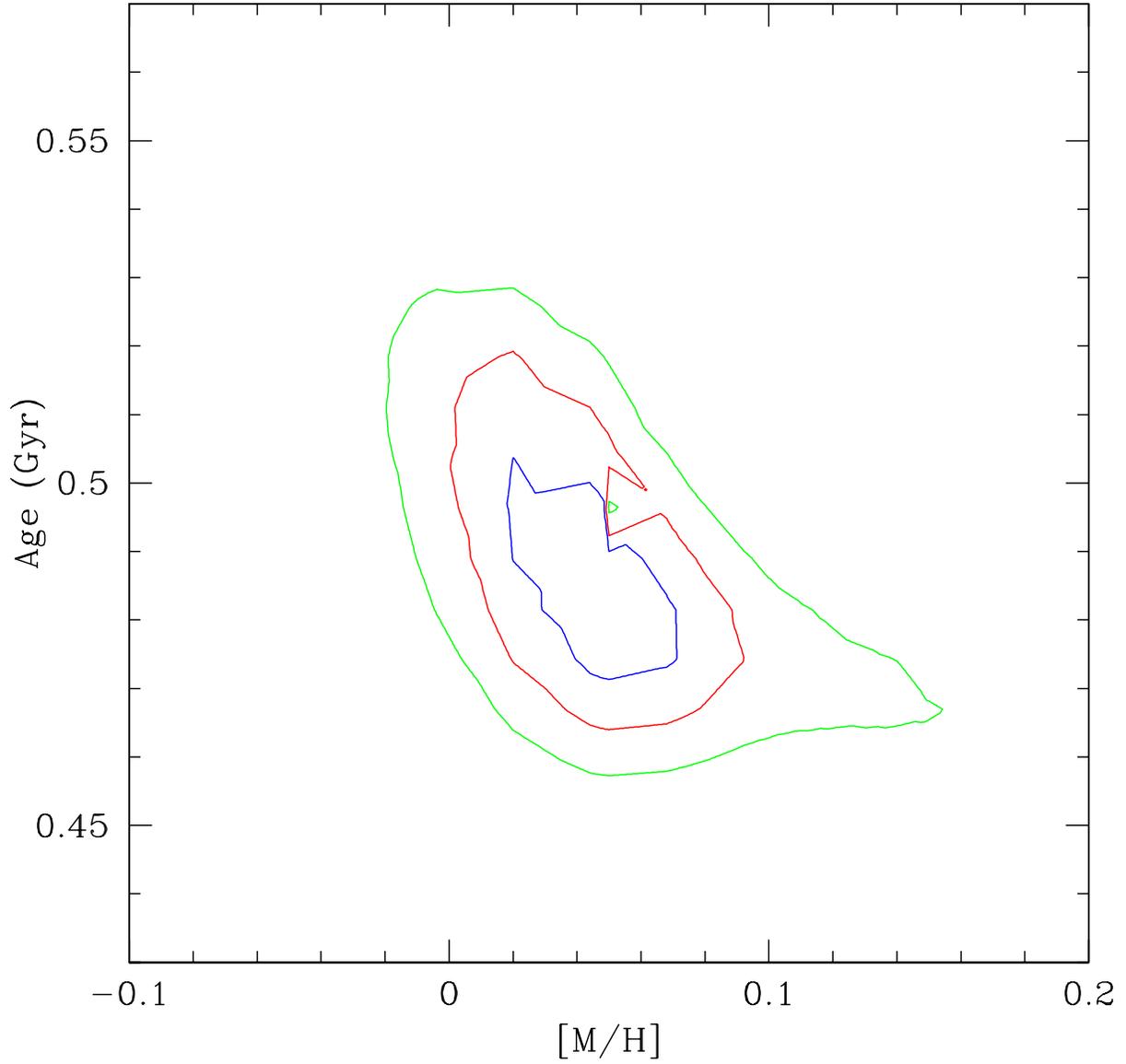}
\caption{Same as figure~\ref{fig:MSfitchi2_1}, in this case we have included spectroscopic constraints on the temperature and metallicity for 12 stars.}
\label{fig:MSfitchi2_2}
\end{figure}

When we include spectroscopic constraints for 12 stars that pass the photometric cuts for being included in the main sequence fit we find: $t = 485 \pm 15$ Myr, $[M/H] = + 0.045 \pm 0.024$, $(m - M)_{V} = 11.572 \pm 0.046$, $E(B-V) = 0.227 \pm 0.008$, $E(V-I_{C}) = 0.355 \pm 0.008$, and $E(V-K_{S}) = 0.695 \pm 0.018$.  We will adopt these parameters for the cluster. Figure~\ref{fig:MSfitchi2_2} shows the resulting age vs. $[M/H]$ $\Delta \chi^{2}$ contour plot. When the spectroscopic constraints are imposed the higher metallicity secondary minimum is rejected. There is one note of caution, the spectroscopic metallicities for the 12 stars show a slight correlation with temperature due to fixing $\log(g)$ in determining $T_{eff}$ and $[M/H]$. However, since the included stars span $\log(g) = 4.5$, the average metallicity of the stars should closely match the true metallicity of the cluster. That the spectroscopic metallicity is in good agreement with the photometric metallicity and the independent spectroscopic determination of $+0.05 \pm 0.05$ by \citet{Marshall.05} suggests that any systematic error is likely to be small.

\placefigure{fig:MSfitchi2_2}

\begin{figure}[p]
\plotone{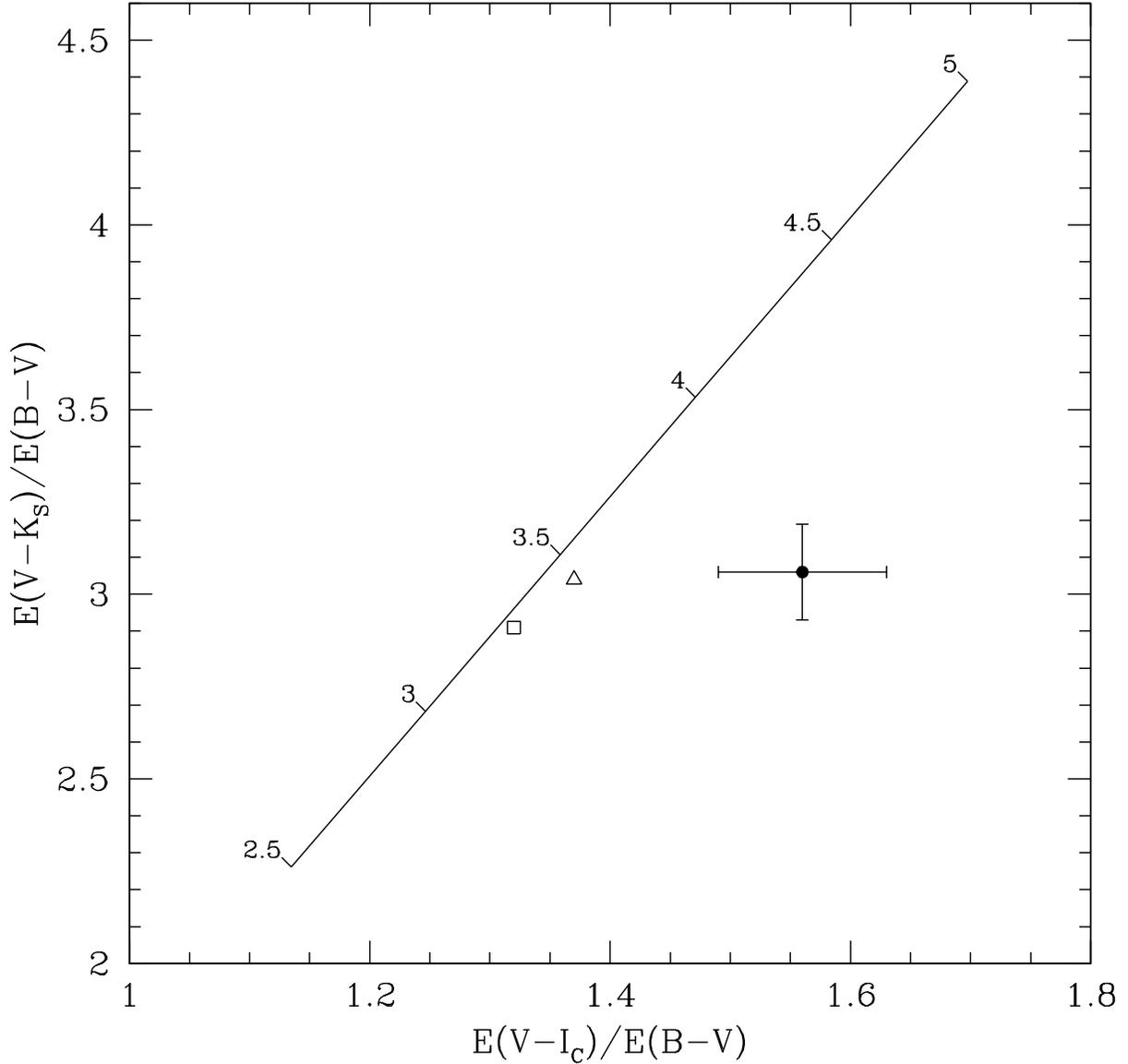}
\caption{Comparison between the observed and theoretical color excess ratios. The observed values (solid point) comes from fitting the YREC isochrones to the main sequence of M37 in $B-V$, $V-I_{C}$ and $V-K_{S}$ CMDs. The line shows the theoretical values from \citet{Cardelli.89} as a function of $R_{V}$ (the numbers plotted along the line). The open square and open triangle correspond to the theoretical values for stars with $(B-V)_{0} = 0$ and $(B-V)_{0} = 0.8$ respectively \citep{Bessell.98} for the \citet{Mathis.90} extinction law. The observed values are inconsistent with the theoretical relation; this may be due to systematic errors in the photometric zero-points (see text).}
\label{fig:EBVEVIEVK_Rv}
\end{figure}

The resulting color excess ratios are: $E(V-I_{C})/E(B-V) = 1.56 \pm 0.07$ and $E(V-K_{S})/E(B-V) = 3.06 \pm 0.13$. For comparison, A07 find $E(V-I_{C})/E(B-V) = 1.32$ and $E(V-K_{S})/E(B-V) = 2.91$ for stars with $(B-V)_{0} = 0.0$ and $E(V-I_{C})/E(B-V) = 1.37$ and $E(V-K_{S})/E(B-V) = 3.04$ for stars with $(B-V)_{0} = 0.8$ using the \citet{Bessell.98} broad-band photometry extinction formulae which are based on the \citet{Mathis.90} extinction law. While the measured $E(V-K_{S})/E(B-V)$ value is in good agreement with the expected value, the measured $E(V-I_{C})/E(B-V)$ value is $13\%$ too high. Varying the ratio of total to selective extinction ($R_{V} = A(V)/E(B-V)$) will not solve this problem as doing so will change the expected value of $E(V-K_{S})/E(B-V)$ in the same sense that $E(V-I_{C})/E(B-V)$ is changed. Figure~\ref{fig:EBVEVIEVK_Rv} illustrates this using the analytic expression for the extinction law given by \citet{Cardelli.89}. We have adopted the effective wavelengths of $\lambda_{eff} = 0.438~\mu{\rm m}$, $0.545~\mu{\rm m}$, and $0.798~\mu{\rm m}$ for the $B$, $V$ and $I_{C}$ filters respectively, which are appropriate for an A0 star \citep{Bessell.98}; for the $K_{S}$ filter we adopt $\lambda_{eff} = 2.159~\mu{\rm m}$ \citep{Cohen.03}. A systematic error of $0.04$ mag in $V-I_{C}$ could account for this discrepancy which is consistent with the error in the transformation from $ri$ to $I_{C}$ together with the systematic error in $I_{C}$ given by \citet{Nilakshi.02}. Note that if we use the Lupton (2005) transformation from $ri$ to $I_{C}$ listed on the SDSS photometric equations website the $V-I_{C}$ values are systematically redder than when we use the transformation to the \citet{Nilakshi.02} system, using this transformation would exacerbate the problem.

\placefigure{fig:EBVEVIEVK_Rv}

The YREC isochrones are generated without core overshooting assuming a non-overshooting age of 550 Myr for the Hyades (see A07). For fitting the main sequence the amount of overshooting has very little effect on the actual shape of the sequence, but will affect the age scale of the measurement. We assume however that the relative age ($t_{M37}/t_{Hyades}$) is independent of the overshooting (where $t_{M37}$ and $t_{Hyades}$ are both measured using the same overshooting). Relative to the Hyades, we estimate that the age is $t_{M37} = (0.88 \pm  0.03) t_{Hyades}$. Note that with moderate overshooting, we'd expect the age to be $551$ Myr using $t_{Hyades} = 625~{\rm Myr}$ as the age of the Hyades with moderate overshooting \citep{Perryman.98}.

\begin{figure}[p]
\plotone{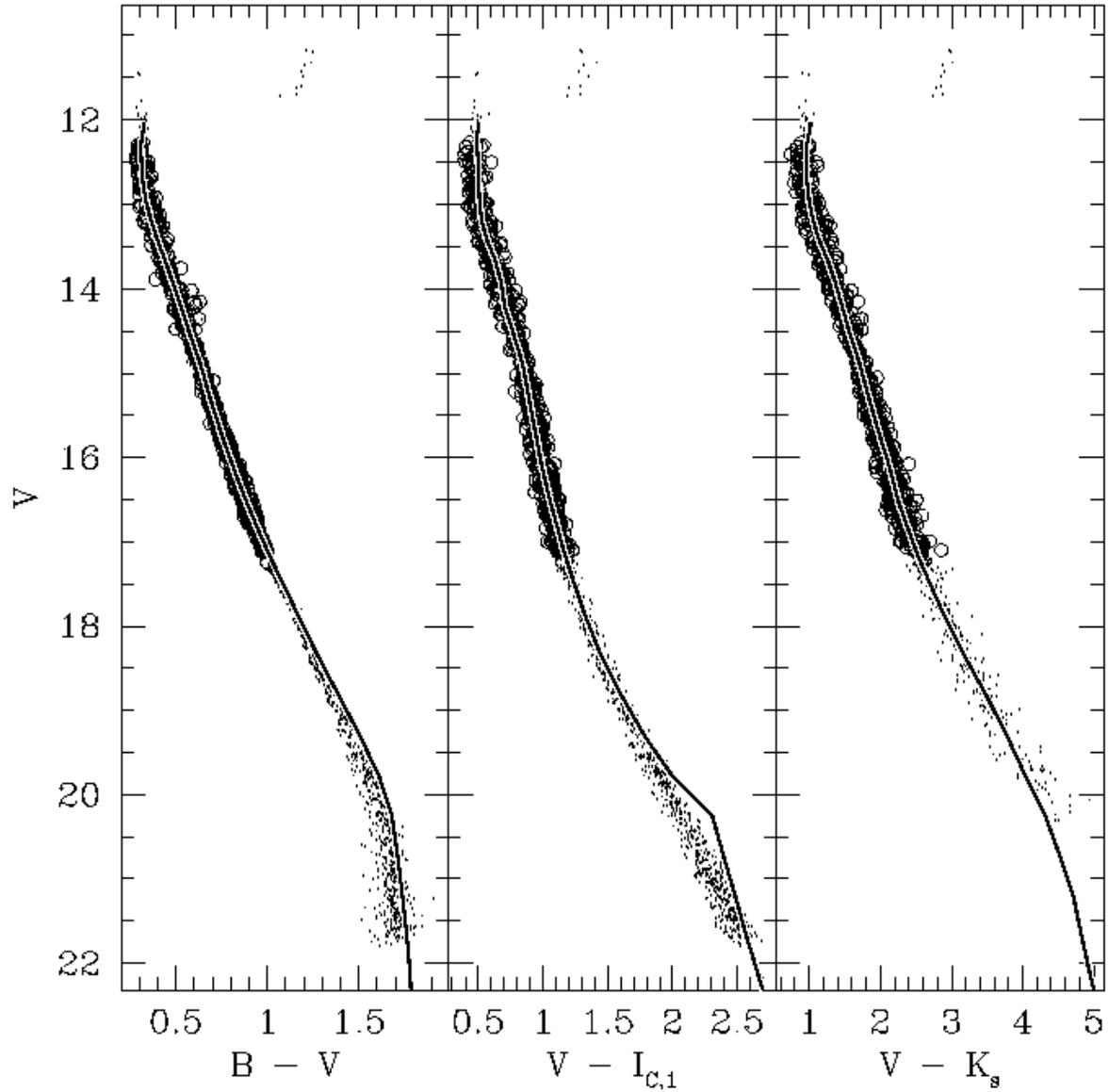}
\caption{Best fit YREC isochrones to the upper portion of the M37 main sequence including spectroscopic constraints for 12 stars. Open circles denote stars that were included in the fit. Note that the isochrones appear to be too red along the lower main sequence. This may be due in part to our neglecting the color dependence of interstellar extinction (redder stars suffer less extinction), but it is also likely due to errors in the model.}
\label{fig:MSfit1}
\end{figure}

Figure~\ref{fig:MSfit1} shows the best fit isochrones, we use open circles to denote the stars that were included in the fit. Note that in this case the lower main sequence stars appear to be noticeably bluer than the isochrones in both the $B-V$ and $V-I_{C}$ CMDs. This is in part due to our neglecting the color dependence of broad-band photometric reddening. For a nominal $E(B-V) = 0.227$ we expect stars with $(B-V)_{0} = 1.0$ to have an $E(B-V)$ value that is $\sim 0.02$ mag smaller than stars with $(B-V)_{0} = 0.0$ \citep{Bessell.98}. However, for stars with $(B-V) > 1.5$, the effect may well be due to errors in the models. One thing to note is that historically theoretical isochrones have been too blue compared to the observed lower main sequence (see for example figures~\ref{fig:Y2isoc} and \ref{fig:Padovaisoc}). The primary difference between the YREC isochrones and other sets is the use of an empirical color-$T_{eff}$ which forces the isochrones to reproduce the Hyades main sequence. According to A07, these empirical relations should be reliable for $(V-I_{C})_{0} < 1.4$. However, we find that there is a discrepancy for stars slightly bluer than this threshold. It is unclear what the source of this discrepancy might be.

\placefigure{fig:MSfit1}

We have also attempted fitting the main sequence down to $(V-I_{C}) < 1.7$. In this case there is a solution with $t = 433 \pm 15$ Myr, $(m-M)_V = 11.79 \pm 0.04$, $E(B-V) = 0.228 \pm 0.009$ and $[Fe/H] = 0.177 \pm 0.011$, or $t = 427 \pm 14$ Myr, $(m-M)_V = 11.81 \pm 0.04$, $E(B-V) = 0.231 \pm 0.010$ and $[Fe/H] = 0.181 \pm 0.014$ when the spectroscopic constraints are not imposed. The errors do not include systematic errors. Because the photometry in this case tends to pull the solution towards a high metallicity and distance modulus that are inconsistent with the spectroscopic measurements, we have chosen to adopt the solution from fitting only the upper main sequence.

\subsubsection{Isochrone Fitting - Y2 and Padova Isochrones}

\begin{figure}[p]
\plotone{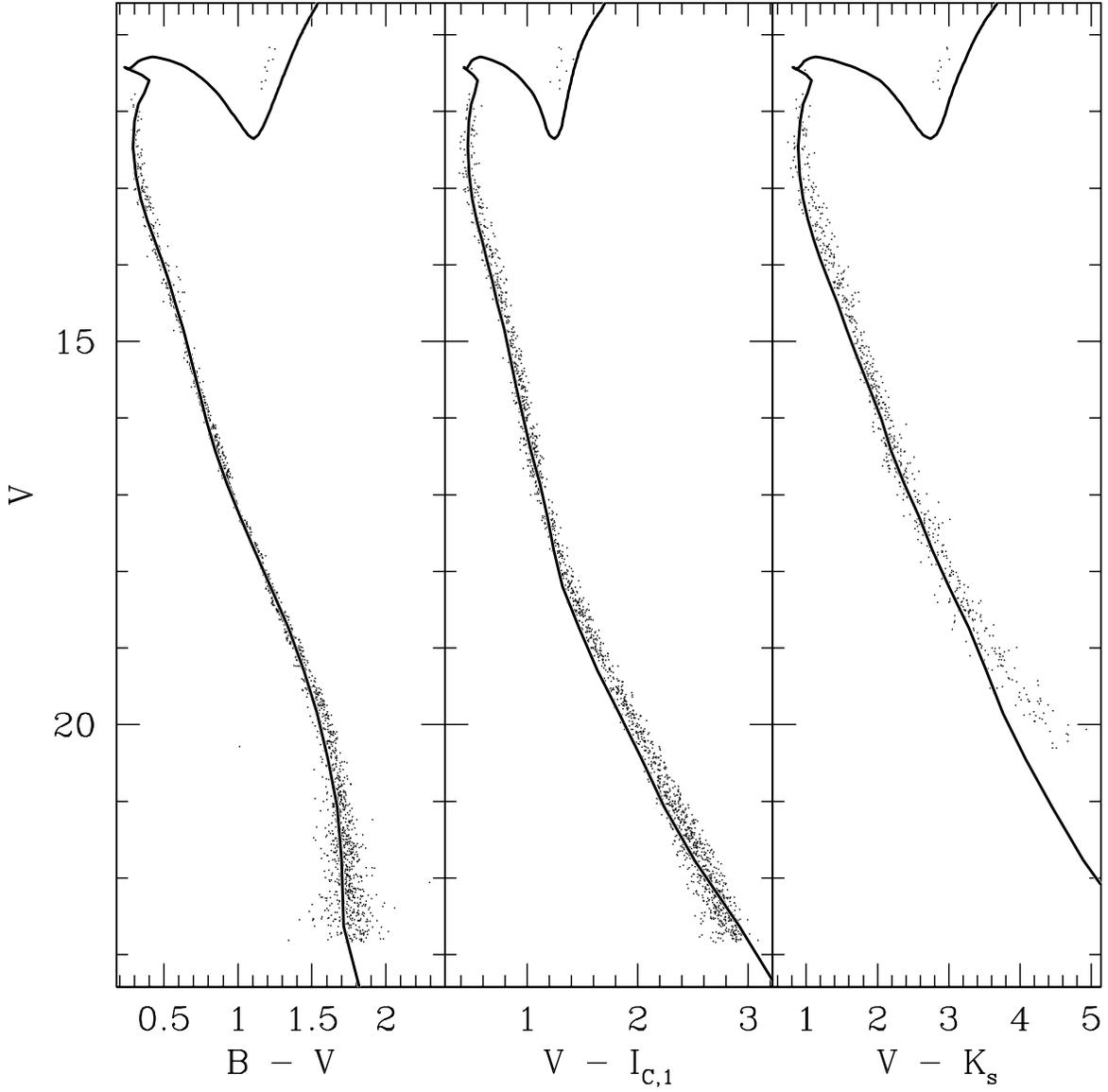}
\caption{$580$ Myr Y2 isochrones plotted on the M37 CMDs. We have assumed $[M/H] = 0.045$, $E(B-V) = 0.227$, $E(V-I_{C}) = 0.355$, $E(V-K_{S}) = 0.695$ and $(m-M)_{V} = 11.572$.}
\label{fig:Y2isoc}
\end{figure}

Since the YREC isochrones do not extend beyond the base of the turnoff, we have explored using other isochrone sets to determine the age of M37. Figure~\ref{fig:Y2isoc} shows the $580$ Myr Yonsei-Yale version 2 core overshooting isochrone \citep[Y2;][]{Demarque.04} plotted on the $B-V$, $V-I_{C}$ and $V-K_{S}$ CMDs. We find that this isochrone, with an estimated uncertainty of $\pm 50$  Myr, best models the turnoff of the cluster. We have assumed the metallicity, distance and reddening from the previous subsection. The Y2 isochrones yield an age for the Hyades of $625$ Myr, thus for this set of isochrones M37 is estimated to have $t = (0.93 \pm 0.08)t_{Hyades}$. The red giant branch (RGB) in figure~\ref{fig:Y2isoc} appears to lie slightly redward of the group of giants, however we note that the the giants are almost certainly red clump, rather than RGB stars. This is expected to be the case since the time a star spends on the red clump is nearly an order of magnitude longer than the time it spends on the RGB. Note that \citet{Kalirai.01} found that the stars are located at the correct position on the blue loop.

\placefigure{fig:Y2isoc}

\begin{figure}[p]
\plotone{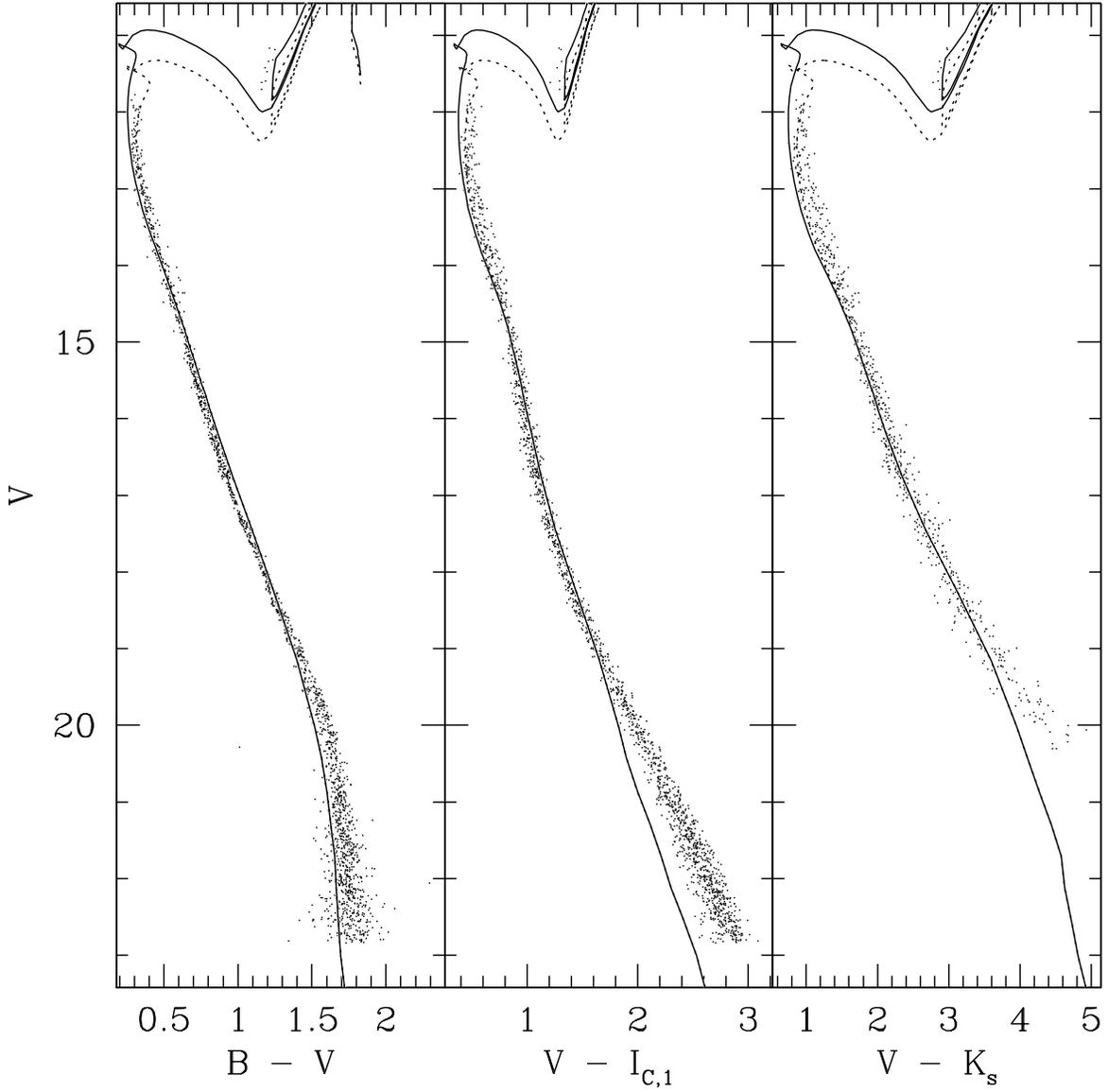}
\caption{Same as figure~\ref{fig:Y2isoc}, here we plot the $540$ Myr Padova isochrones (dotted lines) which fit the main sequence turnoff, and the $430$ Myr Padova isochrones (solid lines) which fit the red clump.}
\label{fig:Padovaisoc}
\end{figure}

We have also attempted to fit the Padova isochrones \citep{Bertelli.94, Girardi.00, Girardi.02} obtained from their web-based interpolator\footnote{http://stev.oapd.inaf.it/\~{}lgirardi/cmd} to the M37 CMDs. Figure~\ref{fig:Padovaisoc} shows the $540$ Myr and $430$ Myr isochrones with solar-scaled abundances. We find that the $540 \pm 50$ Myr isochrone fits the turnoff, while $430 \pm 50$ Myr isochrone fits the red clump. We note, however, that uncertainties in the initial He core mass makes the red clump age estimate less secure than the turnoff estimate. These models do invoke overshooting, but we find that they yield an age of $\sim 575 \pm 25$ Myr for the Hyades. The age of M37 relative to the Hyades using the Padova isochrones is $t = (0.94 \pm 0.1)t_{Hyades}$ when the turnoff is fit or $t = (0.75 \pm 0.09)t_{Hyades}$ when the red clump is fit.

The main sequence turnoff age of M37 relative to the Hyades is consistent among all three isochrone sets. The red clump age from the Padova isochrones, however, is younger than, and marginally inconsistent with the main sequence turnoff age. 

\placefigure{fig:Padovaisoc}

\subsubsection{Isochrone Fitting - Systematic Errors}

It is well known that the errors in the fundamental cluster parameters that result from isochrone fitting are dominated by systematics that can be a factor of 2 to 3 larger than the internal errors from the $\chi^{2}$ fitting. We estimate the systematic uncertainties following the discussion in A07.

Table~\ref{tab:systematics} gives an overview of the sources of systematic error and estimates for the propagated errors on each of the fundamental parameters. The uncertainties on the helium abundance ($Y$) and the errors due to calibration uncertainties in the YREC isochrones are taken from A07. To propagate the helium abundance uncertainty into an uncertainty on the age we use the theoretical relation between age, turnoff luminosity, helium abundance and metallicity from \citet{Iben.84}. To determine the dependence of the parameters on our photometric membership selection threshold, we varied the threshold from $\chi^{2} = 50$ to $\chi^{2} = 250$, the listed errors give the range of parameter values over this interval. We also varied the $B-V$ cut on stars to include in the fit from $B-V < 0.8$ to $B-V < 1.2$. Once again the listed errors correspond to the range of parameter values. The errors listed as $\Delta B$ etc. refer to systematic errors in the photometry.

Our final values for the cluster parameters, including systematic and fitting errors are given in table~\ref{tab:M37finalparams}. The distance of $1490 \pm 120$ pc is calculated assuming $R_{V} = 3.1$.

We note that our systematic errors still do not include the effects of binarity. A07 have conducted numerical simulations which show that for a binary fraction of $50\%$ the photometric metallicity may underestimate the true metallicity by 0.05 dex, the $E(B-V)$ reddening may be overestimated by $0.003$, and the distance modulus may be underestimated by $0.01$.  There is, however, some indication that the binary fraction along the main sequence in this cluster is closer to $20\%$ \citep{Kalirai.04}, so the actual biases will likely be less than this.

\placetable{tab:systematics}

\subsection{Mass/Radius Uncertainty}

For the purposes of the transit survey the most important measurements that come from isochrone fitting are the masses and radii of the main sequence stars. It is therefore necessary to propagate the uncertainties in the fundamental cluster parameters into systematic uncertainties on the masses/radii.

\begin{figure}[p]
\plotone{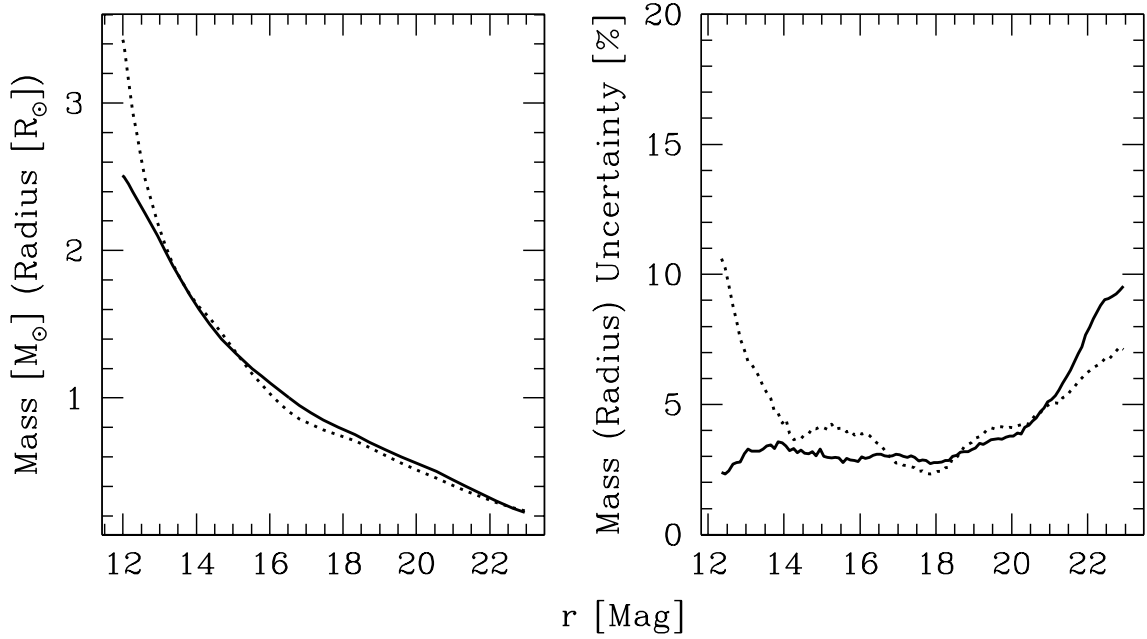}
\caption{Left: The mass (solid line) and radius (dotted line) are plotted as functions of the apparent $r$ magnitude. These relations are calculated from the YREC isochrones assuming the parameters in table~\ref{tab:M37finalparams}. Right: Here we show the percentage systematic uncertainty in mass (solid line) and radius (dotted line) as a function of $r$. The uncertainty is calculated by a Monte Carlo simulation, varying the cluster parameters assuming they follow a normal distribution, it does not account for systematic errors in the isochrones. The uncertainty for stars with $15 < r < 20$ is $\sim 4\%$.}
\label{fig:MRunc}
\end{figure}

In figure~\ref{fig:MRunc} we plot the mass-$r$ and radius-$r$ relations for the cluster calculated from the YREC isochrones assuming the parameters in table~\ref{tab:M37finalparams}. Since all of the stars in our survey have an $r$ measurement, whereas only those overlapping the \citet{Kalirai.01} fields have a $V$ measurement, we first find a fiducial $V-r$ main sequence determined by eye to convert the mass-$V$ and radius-$V$ relations from the isochrones into $r$ magnitudes. The primary time series observations extend from $r \sim 14.5$ (or $M \sim 1.4 M_{\odot}$) down to $r \sim 23$ (or $M \sim 0.2 M_{\odot}$). Figure~\ref{fig:MRunc} also shows the percentage systematic uncertainties on the masses and radii. This is calculated using 1000 Monte Carlo simulations where we assume a normal distribution for each of the cluster parameters and we also assume the errors on the parameters are uncorrelated. Between $15 < r < 20$ the uncertainties on the mass and radius are $\sim 4\%$.

\placefigure{fig:MRunc}

\section{Luminosity and Mass Functions}

In this subsection we use our $gri$ photometry to obtain new estimates for the luminosity and mass functions (LF/MF) of the cluster. Before we do this we must correct our star counts for incompleteness and field star contamination.

To estimate the completeness of our photometric observations we conduct artificial star tests. For each image we conduct ten different tests injecting 100 stars per test. The injected stars have instrumental magnitudes uniformly distributed between 5.0 and 20.0 (corresponding roughly to $9.8 < g < 24.8$, $9.3 < r < 24.3$, and $8.7 < i < 23.7$). We process each simulated image through the DAOPHOT photometry routines described in \S 3.2 and tabulate the recovery frequency as a function of instrumental magnitude for every chip. This is done in each filter for both our shallow and deep observations of the cluster as well as for our observations of the neighboring field.

For every star that we observe we determine a correction factor which is the inverse of the probability of having detected the star in all three filters:
\begin{equation}
c = \frac{1}{f_{g^{\prime}}f_{r^{\prime}}f_{i^{\prime}}}
\end{equation}
where $f = N_{found}/N_{injected}$ is the recovery frequency in the specified filter for the chip used to detect the star and for the instrumental magnitude of the star. For stars that lie in the overlap between the shallow and deep observations we take the detection frequency as the probability that the star would be observed in either the deep or the shallow image $f = f_{deep} + f_{shallow} - f_{deep}f_{shallow}$.

The completeness corrected LF can then be computed in the calibrated magnitude system by summing the correction factors for all stars within in a given magnitude bin. There are two contributions to the errors: uncertainties on the correction factors, and uncertainties on the number of stars. The correction factors are estimated by conducting a number of Bernoulli trials, so the errors on these factors ($\sigma_c$) are just the standard errors on the estimated proportion for a binomial distribution ($\sim c^{2}/\sqrt{N}$ for $N$ injected stars in the given magnitude bin). We assume Poisson errors for the number of detected stars. The error on the LF in a given magnitude bin is then:
\begin{equation}
\sigma_{N}^{2} = n\bar{c}^{2} + \sum_{i=1}^{n}\sigma_{c,i}^{2}
\end{equation}
where the sum is over the $n$ stars that lie within the magnitude bin, and $\bar{c}$ is the average correction factor for these stars.

\begin{figure}[p]
\plotone{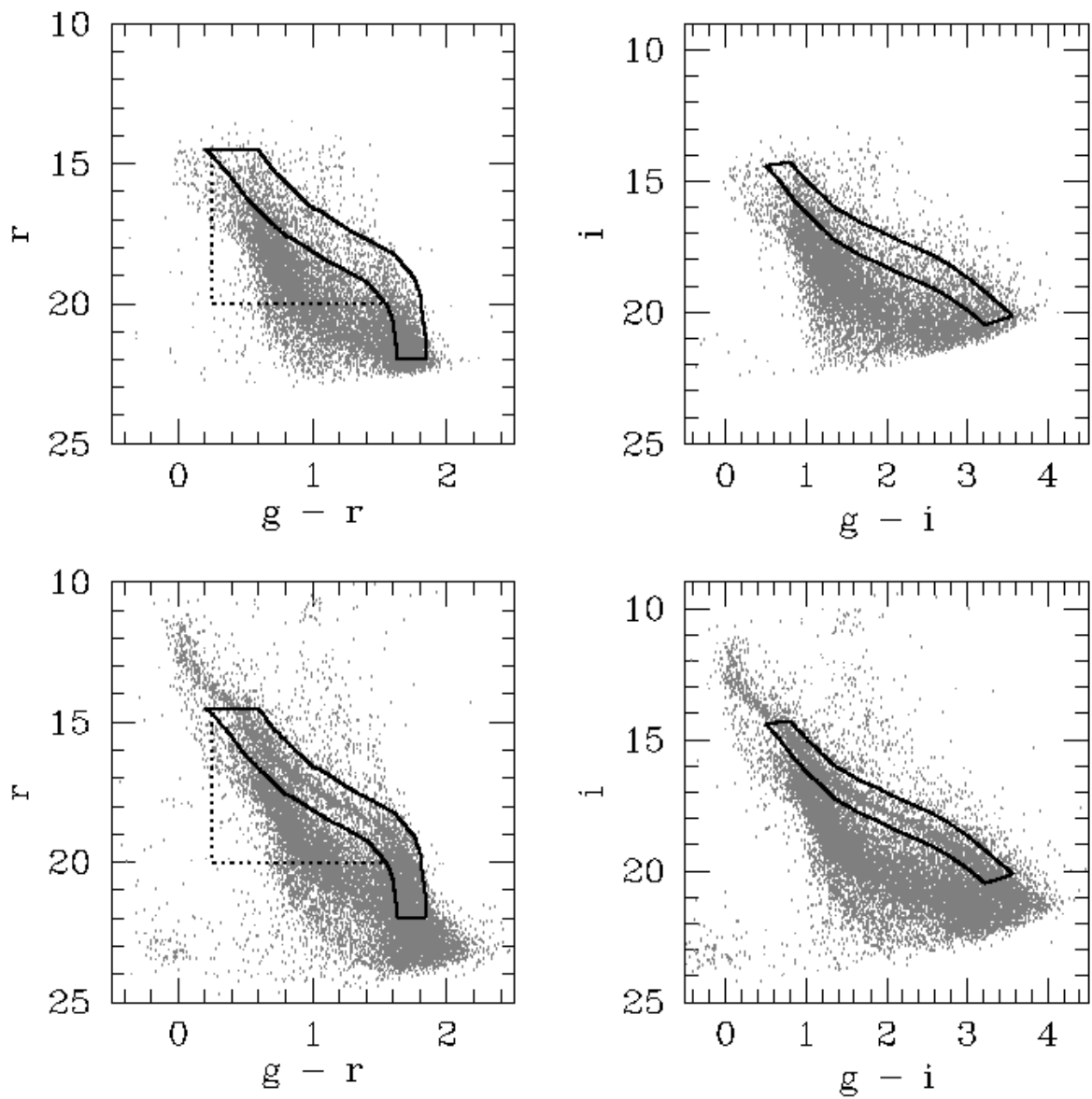}
\caption{The selections of stars on and below the cluster main sequence, which are used in computing the luminosity function (figure~\ref{fig:M37_LF}), are plotted on $g-r$ and $g-i$ CMDs. The top two panels show the CMDs for the field off the cluster, while the bottom two panels show the CMDs for the field on the cluster. Stars falling within the solid black lines on the $g-r$ and $g-i$ CMDs are selected as stars near the cluster main sequence, while stars within the region bounded by the dotted and solid lines on the $g-r$ CMD are selected as stars below the cluster main sequence.}
\label{fig:LFselectiononCMD}
\end{figure}

\begin{figure}[p]
\plotone{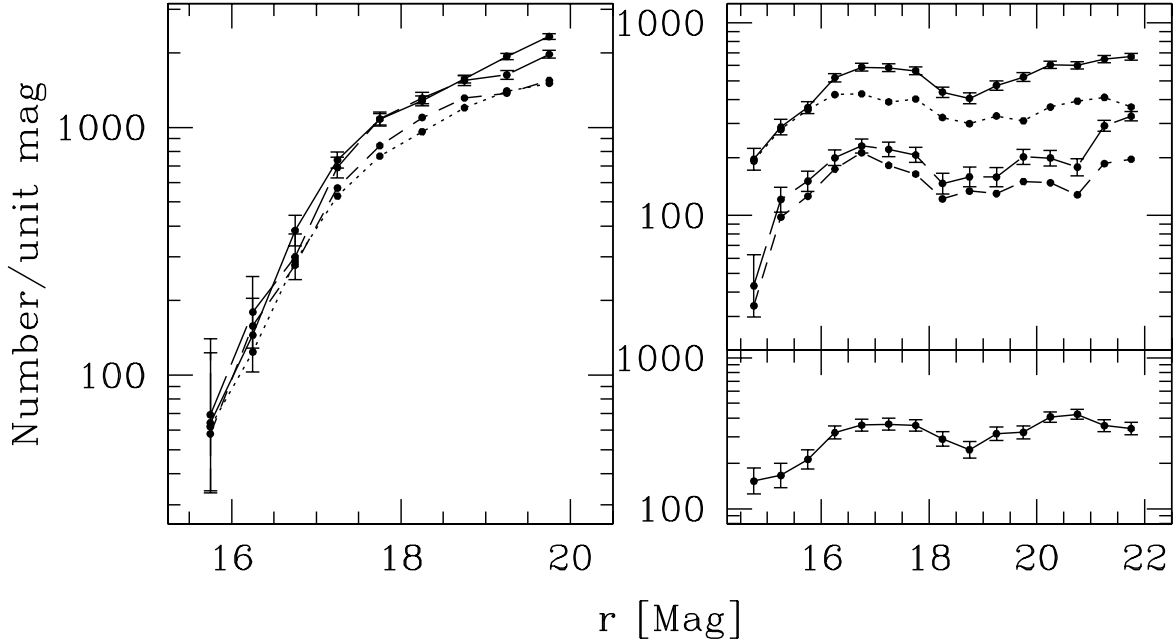}
\caption{Left: The LF for stars below the cluster main sequence on the $g-r$ CMD. The dotted line shows the raw function for the field on the cluster, the short-dashed line shows the raw function for the field off the cluster, the solid line is the completeness corrected function for the field on the cluster and the long-dashed line is the completeness corrected function for the field off the cluster. Note that the completeness corrected functions agree for the on and off-cluster fields down to $r \sim 19$ which shows that the non-cluster population of stars is similar for both fields. The completeness correction is larger in the cluster field because there are more bright stars in this field. Upper Right: The same as for the left panel, this time for stars near the cluster main sequence in the $gri$ CMDs. Bottom Right: The difference between the completeness corrected on-field LF and the completeness corrected off-field LF. This is the completeness and contamination corrected LF for the cluster. The difference in magnitude ranges between the left and right plots is due to the more restrictive range used to select background galactic disk stars that do not intersect the cluster main sequence in the left plot. The faint magnitude limit for the cluster LF is set by the depth of the off-cluster field observations.}
\label{fig:M37_LF}
\end{figure}

We compute the LF for both the on and off-cluster fields for stars located below the main sequence on the $g-r$ CMD and for stars located near the cluster main sequence, these selections are shown in figure~\ref{fig:LFselectiononCMD}. For the latter selection we take the fiducial main sequence isochrones in $g-r$ and $g-i$ and require stars to lie within the region bounded by $(g-r) - 0.1, r + 0.5$ and $(g-r) + 0.1, r - 0.5$ and within the region bounded by $(g-i), i + 0.5$ and $(g-i), i - 0.75$. We use a relatively wide band to select all cluster members, including binaries. This is different from the selection of probable cluster members used in \S 5.1 where we explicitly sought to reject binaries for the purpose of determining the cluster parameters. The resulting LF is shown in figure~\ref{fig:M37_LF}. The agreement for stars below the main sequence between the on and off-cluster fields down to $r \sim 19$ confirms that the population of non-cluster member stars in the off-cluster field is comparable to that in the on-cluster field, we can therefore use this field to perform a contamination correction to our cluster LF. There is a slight discrepancy for sources fainter than $r \sim 19$, so our contamination correction is less reliable at the faint end. Since we do not have short exposure observations of the field off the cluster, we can only compute the contamination corrected LF for $r > 14.5$. The final LF is shown in the lower right panel of figure~\ref{fig:M37_LF}. Over the magnitude range covered by our spectroscopy ($15 < r < 18.6$) we find that the background contamination for stars selected near the main sequence in $gri$ is $38\%$ which is comparable to the contamination of $40\%$ determined from the radial velocity distribution (\S 4.3).

\placefigure{fig:M37_LF}

\placefigure{fig:LFselectiononCMD}

\begin{figure}[p]
\plotone{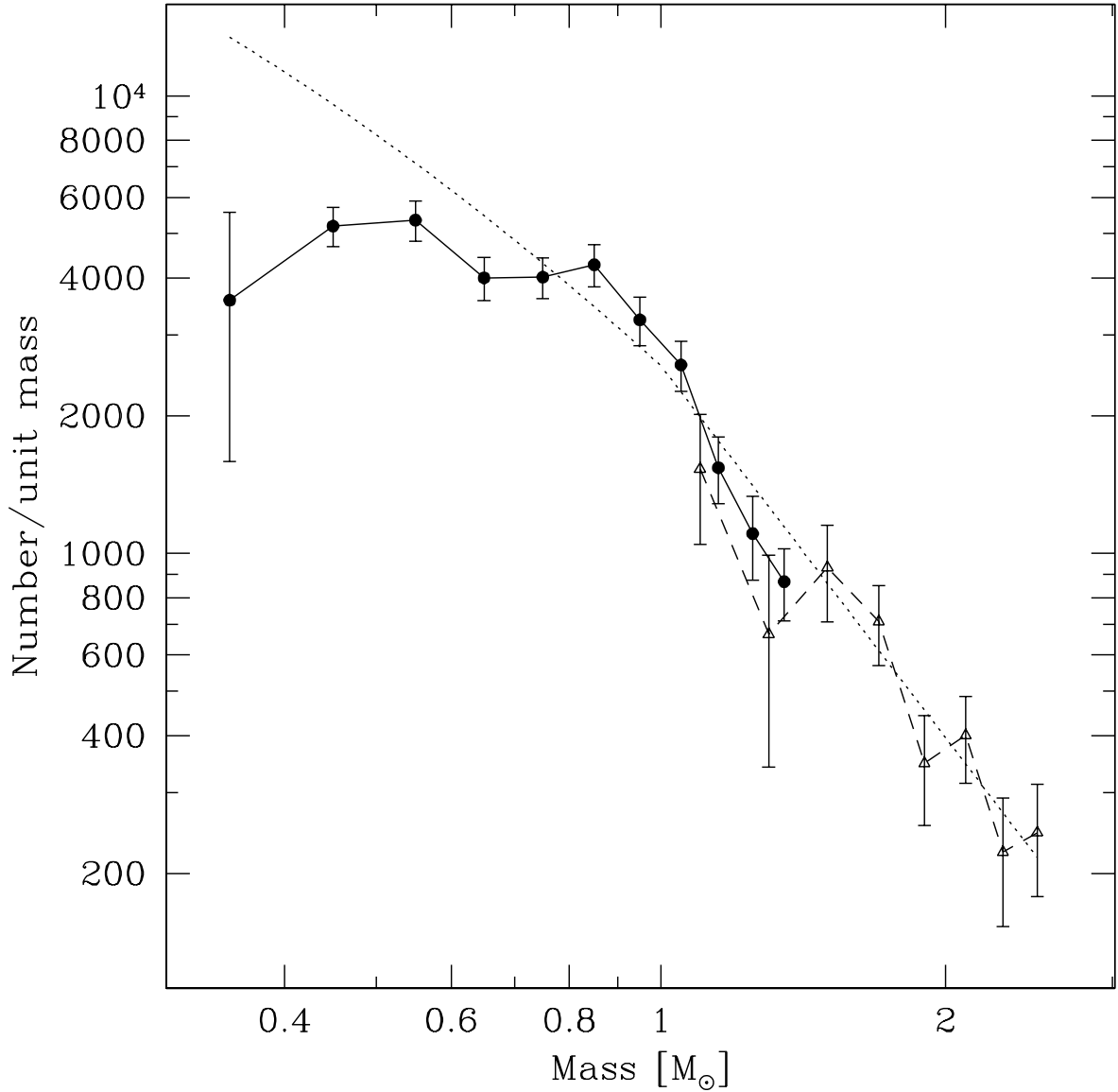}
\caption{Completeness and contamination corrected mass function for M37. The filled points/solid line shows the MF from our $gri$ photometry after correcting for spatial incompleteness. The open triangles/dashed line shows the MF derived from 2MASS photometry. The dotted line shows the galactic IMF determined from field stars and young open clusters (see the text for a description). It has been normalized to match the cluster MF at the high mass end.}
\label{fig:M37_MF}
\end{figure}

To compute the MF we follow a similar procedure as for the LF. In this case we estimate the masses for each star near the cluster main sequence using the mass-$r$ relation described in \S 5.3. Because we do not have very deep or shallow observations of the field off the cluster we use our photometry to compute the MF only over the range $0.3 M_{\odot} < M < 1.4 M_{\odot}$. We also use 2MASS to compute the MF over the range $1.0 M_{\odot} < M < 2.6 M_{\odot}$ using a mass-$K_{S}$ relation derived in a similar fashion to the mass-$r$ relation. The upper mass limit in this case is roughly the turn-off mass of the cluster. We compute the 2MASS MF in bins of width $0.2 M_{\odot}$ for stars within $50\arcmin$ of the cluster center taking the field star density within each mass bin from an annulus of inner radius $50\arcmin$ and width $10\arcmin$. To correct for the spatial incompleteness of our observations we integrate the best fit radial density profile models for the mass bins $0.4 M_{\odot} < M < 0.77 M_{\odot}$, $0.77 M_{\odot} < M < 1.13 M_{\odot}$, $1.13 M_{\odot} < M < 1.5 M_{\odot}$ (see below) over our field of view. We find that our observations have a spatial completeness of $64\%$, $65\%$ and $65\%$ for each of the respective mass bins. We scale our MF in each mass bin by the inverse of the completeness before combining with the 2MASS MF. We assume that stars with $M < 0.4 M_{\odot}$ have a spatial completeness of $64\%$. The resulting completeness and contamination corrected MF for the cluster is shown in figure~\ref{fig:M37_MF} with the data given in tables~\ref{tab:M37_MF}~and~\ref{tab:M37_MF_2MASS}.

As reviewed recently by \citet{Chabrier.03b}, the initial mass function (IMF) for the galactic field and young open clusters follows a power law of the form:
\begin{equation}
\Psi (M) = \Psi_{0} M^{-(1+x)}
\end{equation}
for massive stars, $M \ga 1.0 M_{\odot}$, with an exponent of $x = 1.7$ \citep{Scalo.86}. For lower mass stars the IMF appears to follow a log-normal form \citep{Chabrier.03a}:
\begin{equation}
\Psi (M) \propto M^{-1}\exp [ -\frac{(\log M - \log 0.22)^2}{2 \times (0.57)^2}].
\end{equation}
Both cases include unresolved binaries. The dotted line in figure~\ref{fig:M37_MF} shows this IMF after normalizing to match the high mass cluster data. For $M \ga 0.8 M_{\odot}$ the observed cluster MF is in good agreement with the galactic IMF. Below this value, however, the cluster MF flattens out. This indicates that dynamical processes such as mass segregation and evaporation are important for this cluster. This is expected given that the cluster is slightly older than its relaxation time \citep{Kalirai.01}. 

\placefigure{fig:M37_MF}

Integrating the mass function we estimate that the cluster contains $3960 \pm 240$ stars with $0.3 M_{\odot} < M < 2.6 M_{\odot}$ which contribute $3360 M_{\odot} \pm 150 M_{\odot}$ to the total mass. Extrapolating to the hydrogen burning limit assuming a flat mass function in $M$ as seen in figure~\ref{fig:M37_MF}, we estimate that there are $\sim 790 \pm 440$ stars with $0.08 M_{\odot} < M < 0.3 M_{\odot}$ which contribute $150 M_{\odot} \pm 80 M_{\odot}$ to the cluster mass. \citet{Mermilliod.96} have shown that the cluster contains 35 red giants, which, assuming a mass of $\sim 2.5 M_{\odot}$, would contribute $90 M_{\odot}$ to the cluster mass. Finally \citet{Kalirai.05} estimate that the cluster contains $\sim 50$ white dwarfs with masses between $0.7 - 0.9 M_{\odot}$. So we conclude that the cluster contains $\sim 4800$ stars down to the hydrogen burning limit and has a total mass of $3640 M_{\odot} \pm 170 M_{\odot}$. This is somewhat larger than the 4000 stars and $\sim 2500 M_{\odot}$ determined by \citet{Kalirai.01}; the discrepancy is most likely due to their uncorrected slight spatial incompleteness (their field of view was $42\arcmin x 28\arcmin$), as well as to differences in the estimated slope of the MF for $\sim 1 M_{\odot}$ stars.

Of particular interest for the transit survey is the number of cluster members for which we have obtained time series photometry. We estimate that this number is $1610 \pm 90$ within the mass range $0.3 M_{\odot} < M < 1.4 M_{\odot}$ (corresponding to the magnitude range $14.69 < r < 22.18$).

\section{Radial Density Profile}

We will now investigate the radial density distribution of the cluster which will give us information about its dynamical state. Both \citet{Kalirai.01} and \citet{Nilakshi.02} have previously determined the radial density profile using their own photometry. Because our field of view does not extend to the tidal radius of the cluster we will first measure the tidal radius using 2MASS. The advantage of using 2MASS is that one can extract uniform photometry over a wide field that extends beyond the tidal radius. The average field star density is therefore more secure than the determinations by \citet{Kalirai.01} and \citet{Nilakshi.02}. The disadvantage of 2MASS is that it does not go nearly as deep as the optical surveys so it cannot be used to investigate processes such as mass segregation.

\begin{figure}[p]
\plotone{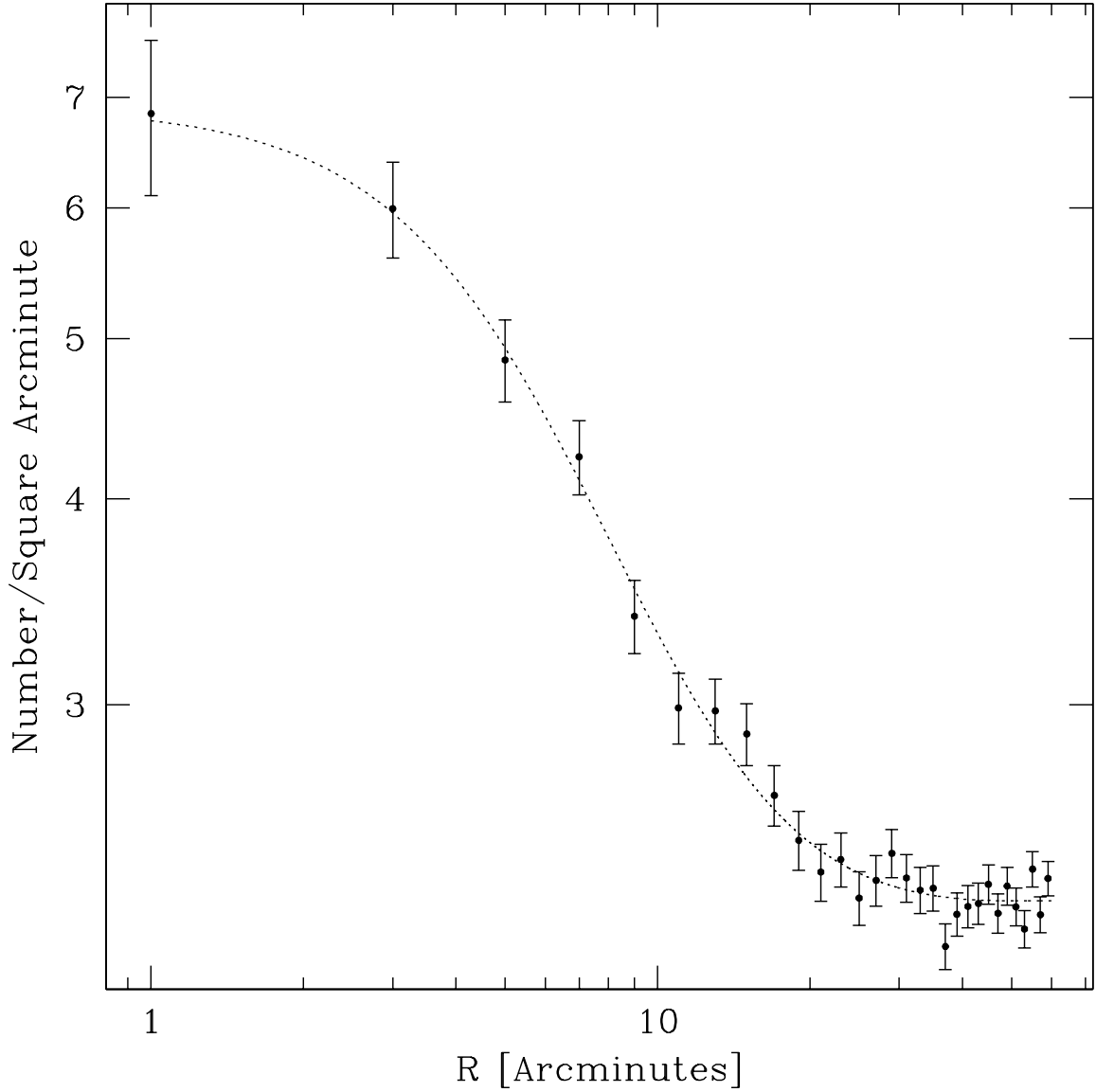}
\caption{The radial density distribution of 2MASS stars with $8.0 < K_{S} < 15.0$. The data has not been corrected for field star contamination. The dotted line shows the best fit King profile which has a tidal radius of $r_{t} = 50\arcmin \pm 15\arcmin$ and a core radius of $r_{c} = 6\farcm4 \pm 0\farcm8$. An average field star density is included in the model.}
\label{fig:M37_2MASS_radialdensity}
\end{figure}

Figure~\ref{fig:M37_2MASS_radialdensity} shows the radial density distribution of stars from 2MASS in bins of width $2\arcmin$. We have included all stars within the range $8.0 < K_{S} < 15.0$ to avoid completeness problems at the faint end, and have adopted the cluster center of 05:52:17.6, 32:32:08 (J2000) from \citet{Kalirai.01}. They estimate that the error on the center is $\pm 30\arcsec$ and is dominated by small number statistics. The data plotted in this figure does not have the field star contribution subtracted. The dotted line shows the best fit single-mass King profile \citep{King.62} which has a tidal radius of $r_{t} = 50\arcmin \pm 15\arcmin$, and a core radius of $r_{c} = 6\farcm4 \pm 0\farcm8$. The other two free parameters in the fit are the central density of $k_c = 6.0 \pm 0.6$ stars/square arcminute and the field density of $k_f = 2.28 \pm 0.02$ stars/square arcminute. Of these four parameters only $r_{t}$ should be insensitive to the specific magnitude range and filter used.  

Using the distance determined in \S 5.2, we find that the tidal radius is $r_{t} = 22 \pm 7$ pc which is in agreement with the expected value of 22.8 pc \citep[we have recalculated the expected value using our new mass estimate]{Kalirai.01}, assuming a mass of the Galaxy within the cluster's orbit of $1.02\times 10^{11} M_{\odot}$ \citet{Clemens.85} and a Galactocentric distance of $10$ kpc. 

\placefigure{fig:M37_2MASS_radialdensity}

\begin{figure}[p]
\plotone{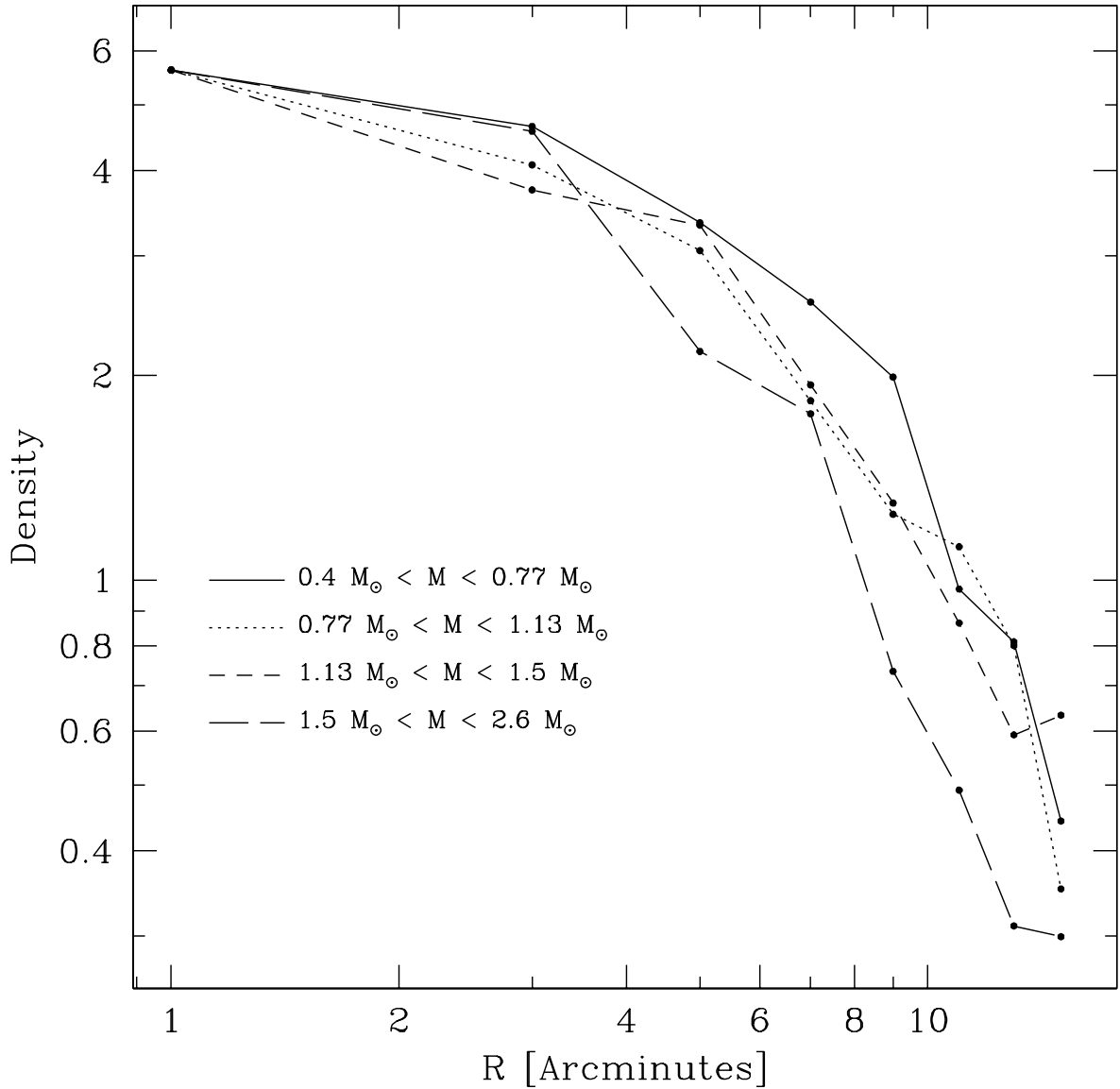}
\caption{The radial density distribution is determined for four mass bins. We have subtracted from each curve the average star density for the relevant mass bin measured in the off-cluster field, and have scaled the curves to have the same central density. The data for the highest mass bin comes from 2MASS for which the background density was determined in an annulus surrounding the cluster (see the text). Note that the units on the vertical scale are arbitrary. The higher mass stars appear to be more centrally concentrated which indicates that the cluster does exhibit mass segregation.}
\label{fig:M37_radialdensity_massbins}
\end{figure}

In figure~\ref{fig:M37_radialdensity_massbins} we show the radial density distribution corrected for field star contamination in four different mass bins. The stars are selected into the first three mass bins based on their proximity to the cluster main sequence in the $g-r$ and $g-i$ CMDs, whereas the fourth mass bin consists of 2MASS detections. We have subtracted an average star density from each bin measured using the off-cluster observations, and have rescaled the curves to have the central density of the lowest mass bin. The higher mass stars appear to be more centrally concentrated than the lower mass stars. To quantify this we fit a King Model to each of the mass bins fixing the tidal radius to the value found above and the field density to the values measured using the off-cluster observations. We find $r_{c} = 6\farcm6 \pm 0\farcm6$ for $0.4 M_{\odot} < M < 0.77 M_{\odot}$, $r_{c} = 6\farcm3 \pm 0\farcm5$ for $0.77 M_{\odot} < M < 1.13 M_{\odot}$, $r_{c} = 6\farcm1 \pm 0\farcm4$ for $1.13 M_{\odot} < M < 1.5 M_{\odot}$ and $r_{c} = 4\farcm0 \pm 0\farcm4$ for $1.5 M_{\odot} < M < 2.6 M_{\odot}$. Fitting a linear relation between $M$ and $r_{c}$ we find $r_{c} \propto (-1.9 \pm 0.4)M$ so that the probability that the system does not exhibit mass segregation (i.e. the probability that the slope in the above relation is consistent with zero or positive) is less than $1\times10^{-5}$.

\placefigure{fig:M37_radialdensity_massbins}

\section{Conclusion}

We have introduced a deep transit survey of the open cluster M37. This survey is unique among open cluster variability surveys, and among time-series surveys in general, in terms of the number of images acquired with such a large telescope. This effectively allows us to study low-amplitude variability among low mass stars in a richer and older cluster than can be targeted with a smaller telescope. We therefore expect to have a unique data set for studying general stellar variability, stellar rotation and small amplitude transiting planets. We will discuss each of these topics in future papers.

In this paper we have laid the groundwork for these future investigations by refining the fundamental cluster parameters. By comparing photometry and high resolution spectroscopy with an empirically calibrated set of theoretical main sequence isochrones we have determined $t = 485 \pm 28$ Myr without overshooting, $[M/H] = +0.045 \pm 0.044$, $E(B-V) = 0.227 \pm 0.038$ and $(m-M)_{V} = 11.572 \pm 0.13$ which are in good agreement with, though more precise than, previous measurements. We have obtained mass and luminosity functions and have shown that they are consistent with the Galactic IMF down to $0.8 M_{\odot}$ with evidence for evaporation below that level. We have also found mass segregation by comparing the radial density profiles in different mass bins.

\acknowledgements
This research has made use of the SIMBAD database, operated at CDS, Strasbourg, France. We would like to thank the anonymous referee for suggestions which improved the quality of this paper. We are grateful to C.~Alcock for providing partial support for this project through his NSF grant (AST-0501681). Funding for M.~Holman came from NASA Origins grant NNG08GH69G. We wish to thank G.~F\"{u}r\'{e}sz and A.~Szentgyorgyi for help in preparing the Hectochelle observations, G.~Torres for help in planning the observations and helpful comments on this paper, and the staff of the MMT, without whom this work would not have been possible. We would also like to thank the TAC for awarding a significant amount of telescope time for this project.

\clearpage

\begin{deluxetable}{lrrr}
\tabletypesize{\scriptsize}
\tablewidth{0pc}
\tablecaption{Summary of Photometric Observations}
\tablehead{\colhead{Date} & \colhead{FWHM($\arcsec$)\tablenotemark{a}} & \colhead{RMS FWHM($\arcsec$)} & \colhead{\# Images/EXPTIME (sec)\tablenotemark{b}}}
\startdata
\cutinhead{On cluster field - $g^{\prime}$ filter}
12/23/2005 & 0.88 & 0.05 & 1/30, 4/60\\
12/24/2005 & 1.12 & 0.36 & 5/1, 15/30\\
12/27/2005 & 1.60 & 0.05 & 5/60\\
12/30/2005 & 1.42 & 0.36 & 5/2, 5/70\\
Total      & 1.18 & 0.36 & 10/1-2, 30/30-70\\
\cutinhead{On cluster field - $r^{\prime}$ filter}
12/21/2005 & 1.19 & 0.24 & 252/50-77\\
12/22/2005 & 0.68 & 0.15 & 404/30-90\\
12/23/2005 & 0.75 & 0.09 & 4/15, 238/30-300\\
12/24/2005 & 1.12 & 0.42 & 5/1, 239/30-140\\
12/26/2005 & 0.96 & 0.28 & 226/35-170\\
12/27/2005 & 1.42 & 0.49 & 2/10, 134/60-360\\
12/28/2005 & 0.85 & 0.19 & 376/30-120\\
12/29/2005 & 0.82 & 0.32 & 330/30-120\\
12/30/2005 & 1.17 & 0.30 & 5/2, 202/40-160\\
12/31/2005 & 1.01 & 0.28 & 8/20, 215/40-120\\
01/01/2006 & 2.00 & 0.30 &  29/30-120\\
01/02/2006 & 0.97 & 0.28 & 193/40-120\\
01/03/2006 & 1.18 & 0.16 &  1/5, 60/70-120\\
01/04/2006 & 2.67 & 0.40 &  2/5, 28/60-120\\
01/05/2006 & 1.11 & 0.31 & 1/1, 8/5, 2/10, 124/45-150\\
01/06/2006 & 2.20 & 0.42 &  1/5, 86/90-120\\
01/07/2006 & 1.01 & 0.19 & 1/5, 2/10, 301/30-90\\
01/08/2006 & 1.32 & 0.35 &  4/5, 82/60-90\\
01/09/2006 & 0.88 & 0.20 & 380/40-90\\
01/10/2006 & 0.69 & 0.15 & 1/10, 445/30-60\\
01/18/2006 & 0.66 & 0.23 & 266/30-90\\
01/19/2006 & 0.80 & 0.14 & 102/50-160\\
01/20/2006 & 1.35 & 0.30 &  47/30-150\\
01/21/2006 & 0.99 & 0.25 & 156/40-200\\
Total      & 0.89 & 0.39 & 47/1-20, 4916/30-360\\
\cutinhead{On cluster field - $i^{\prime}$ filter}
12/24/2005 & 0.65 & 0.07 & 5/0.5, 15/30\\
12/27/2005 & 1.57 & 0.07 & 5/60\\
12/30/2005 & 1.14 & 0.14 & 5/2, 5/70\\ 
Total      & 0.76 & 0.35 & 10/0.5-2, 25/30-70\\
\cutinhead{Off cluster field - $g^{\prime}$ filter}
12/30/2005 & 1.44 & 0.01 & 5/70\\
01/07/2006 & 0.96 & 0.04 & 5/60\\
Total      & 1.23 & 0.24 & 10/60-70\\
\cutinhead{Off cluster field - $r^{\prime}$ filter}
12/30/2005 & 1.33 & 0.02 & 5/70\\
01/07/2006 & 0.90 & 0.02 & 5/60\\
Total      & 1.11 & 0.22 & 10/60-70\\
\cutinhead{Off cluster field - $i^{\prime}$ filter}
12/30/2005 & 1.28 & 0.15 & 16/70\\
01/07/2006 & 0.86 & 0.03 & 5/60\\
Total      & 1.25 & 0.23 & 10/60-70\\
\cutinhead{SDSS Field 1 - $g^{\prime}$ filter}
12/24/2005 & 1.87 & 0.12 & 5/10\\
12/27/2005 & 1.87 & 0.38 & 10/10\\
Total      & 1.87 & 0.32 & 15/10\\
\cutinhead{SDSS Field 1 - $r^{\prime}$ filter}
12/24/2005 & 1.37 & 0.13 & 5/10\\
12/27/2005 & 1.67 & 0.10 & 10/10\\
Total      & 1.61 & 0.16 & 15/10\\
\cutinhead{SDSS Field 1 - $i^{\prime}$ filter}
12/24/2005 & 1.39 & 0.37 & 5/10\\
12/27/2005 & 1.44 & 0.34 & 10/10\\
Total      & 1.39 & 0.35 & 15/10\\
\cutinhead{SDSS Field 2 - $g^{\prime}$ filter}
12/28/2005 & 1.19 & 0.16 & 30/10\\
12/30/2005 & 0.88 & 0.06 & 5/10\\
01/05/2006 & 1.57 & 0.22 & 10/10\\
Total      & 1.22 & 0.25 & 45/10\\
\cutinhead{SDSS Field 2 - $r^{\prime}$ filter}
12/28/2005 & 1.01 & 0.12 & 25/10\\
12/30/2005 & 0.71 & 0.05 & 5/10\\
01/05/2006 & 1.48 & 0.30 & 10/10\\
Total      & 1.04 & 0.30 & 40/10\\
\cutinhead{SDSS Field 2 - $i^{\prime}$ filter}
12/30/2005 & 0.80 & 0.10 & 5/10\\
\enddata
\tablenotetext{a}{Median over images on that date.}
\tablenotetext{b}{For the $r^{\prime}$ time-series images of M37 we give the range of exposure times used.}
\label{tab:obssummary}
\end{deluxetable}

\clearpage

\begin{deluxetable}{rr}
\tabletypesize{\small}
\tablewidth{0pc}
\tablecaption{Exposure Time as a Function of Seeing}
\tablehead{\colhead{Seeing($\arcsec$)} & \colhead{Exposure Time (seconds)}}
\startdata
$< .65$ & 30\\
0.8 & 45\\
1.0 & 60\\
1.1 & 68\\
1.3 & 72\\
\enddata
\label{tab:exptimes}
\end{deluxetable}

\clearpage

\begin{deluxetable}{lr}
\tabletypesize{\small}
\tablewidth{0pc}
\tablecaption{Summary of Hectochelle Spectroscopic Observations of M37}
\tablehead{\colhead{Date} & \colhead{Number of Exposures}}
\startdata
02/23/2007 & 9\\
03/03/2007 & 5\\
03/04/2007 & 4\\
03/06/2007 & 4\\
03/11/2007 & 7\\
03/12/2007 & 6\\
\hline
Total & 35\\
\enddata
\label{tab:specsummary}
\end{deluxetable}

\clearpage

\begin{deluxetable}{lrrrrrr}
\tabletypesize{\scriptsize}
\tablewidth{0pc}
\tablecaption{Zero-points in the transformation from instrumental to Sloan magnitudes.}
\tablehead{\colhead{Chip} & \colhead{$z_{g,j}$} & \colhead{Err $z_{g,j}$} & \colhead{$z_{r,j}$} & \colhead{Err $z_{r,j}$} & \colhead{$z_{i,j}$} & \colhead{Err $z_{i,j}$}}
\startdata
 1 & -0.012 &  0.008 & -0.011 &  0.005 &  0.025 &  0.007 \\
 2 & -0.066 &  0.009 &  0.001 &  0.005 &  0.031 &  0.006 \\
 3 & -0.033 &  0.009 &  0.011 &  0.005 & -0.039 &  0.006 \\
 4 &  0.043 &  0.007 &  0.011 &  0.004 & -0.037 &  0.006 \\
 5 &  0.015 &  0.009 & -0.009 &  0.005 & -0.026 &  0.007 \\
 6 &  0.048 &  0.010 & -0.008 &  0.005 & -0.030 &  0.006 \\
 7 & -0.051 &  0.008 & -0.016 &  0.004 & -0.001 &  0.006 \\
 8 & -0.051 &  0.008 & -0.021 &  0.004 & -0.001 &  0.006 \\
 9 & -0.061 &  0.009 & -0.031 &  0.004 &  0.006 &  0.006 \\
10 & -0.008 &  0.008 &  0.005 &  0.005 &  0.020 &  0.007 \\
11 &  0.046 &  0.007 &  0.010 &  0.004 &  0.032 &  0.006 \\
12 &  0.087 &  0.009 &  0.035 &  0.005 &  0.038 &  0.006 \\
13 &  0.080 &  0.008 &  0.059 &  0.004 &  0.034 &  0.006 \\
14 &  0.086 &  0.007 &  0.043 &  0.004 & -0.026 &  0.006 \\
15 &  0.078 &  0.008 &  0.012 &  0.005 & -0.037 &  0.006 \\
16 &  0.091 &  0.009 &  0.013 &  0.005 & -0.048 &  0.006 \\
17 &  0.002 &  0.008 & -0.009 &  0.004 & -0.004 &  0.006 \\
18 & -0.012 &  0.008 & -0.017 &  0.004 &  0.014 &  0.006 \\
19 & -0.038 &  0.008 &  0.000 &  0.005 &  0.021 &  0.006 \\
20 &  0.002 &  0.009 &  0.009 &  0.005 &  0.029 &  0.006 \\
21 &  0.070 &  0.015 &  0.037 &  0.006 &  0.036 &  0.007 \\
22 &  0.063 &  0.007 &  0.052 &  0.004 &  0.021 &  0.006 \\
23 &  0.072 &  0.009 &  0.021 &  0.005 &  0.008 &  0.007 \\
24 &  0.047 &  0.007 &  0.001 &  0.004 & -0.015 &  0.006 \\
25 &  0.014 &  0.008 & -0.004 &  0.004 & -0.023 &  0.006 \\
26 & -0.011 &  0.008 & -0.020 &  0.004 &  0.005 &  0.006 \\
27 & -0.064 &  0.007 & -0.017 &  0.004 &  0.017 &  0.007 \\
28 & -0.080 &  0.008 &  0.002 &  0.004 &  0.020 &  0.007 \\
29 & -0.071 &  0.007 & -0.003 &  0.004 &  0.018 &  0.006 \\
30 & -0.027 &  0.008 &  0.005 &  0.004 &  0.024 &  0.006 \\
31 & -0.034 &  0.008 & -0.018 &  0.004 &  0.031 &  0.006 \\
32 &  0.015 &  0.007 & -0.015 &  0.004 & -0.031 &  0.006 \\
33 & -0.003 &  0.008 & -0.029 &  0.005 & -0.051 &  0.006 \\
34 & -0.028 &  0.008 & -0.026 &  0.004 & -0.055 &  0.006 \\
35 & -0.084 &  0.008 & -0.030 &  0.005 & -0.002 &  0.006 \\
36 & -0.125 &  0.009 & -0.043 &  0.005 & -0.003 &  0.007 \\
\enddata
\label{tab:photcalparams}
\end{deluxetable}

\clearpage

\begin{deluxetable}{lrrrr}
\tabletypesize{\footnotesize}
\tablewidth{0pc}
\tablecaption{Airmass and color coefficients in the transformation from instrumental to Sloan magnitudes.}
\tablehead{\colhead{Filter} & \colhead{$a$} & \colhead{Err $a$} & \colhead{$b$} & \colhead{Err $b$}}
\startdata
$g$ & 0.148 & 0.004 & -0.122 & 0.002 \\
$r$ & 0.073 & 0.002 & -0.107 & 0.001 \\
$i$ & 0.037 & 0.003 & -0.137 & 0.002 \\
\enddata
\label{tab:photcalparams2}
\end{deluxetable}

\clearpage

\begin{deluxetable}{rrrrrrrrrr}
\tabletypesize{\footnotesize}
\tablewidth{0pc}
\tablecaption{Photometric Catalog of M37\tablenotemark{a}}
\tablehead{\colhead{ID\tablenotemark{b}} & \colhead{RA (J2000)} & \colhead{DEC (J2000)} & \colhead{$g$} & \colhead{$r$} & \colhead{$i$} & \colhead{$\sigma_{g}$} & \colhead{$\sigma_{r}$} & \colhead{$\sigma_{i}$}}
\startdata
010001 & 05:53:03.65 & +32:45:18.1 & 15.383 & 14.789 & 14.491 &  0.008 &  0.007 &  0.008 \\
010002 & 05:52:50.72 & +32:44:32.4 & 15.323 & 14.860 & 14.621 &  0.008 &  0.007 &  0.008 \\
010003 & 05:53:16.91 & +32:45:05.1 & 15.377 & 14.873 & 14.608 &  0.008 &  0.007 &  0.008 \\
010004 & 05:52:54.51 & +32:44:52.3 & 15.434 & 14.956 & 14.688 &  0.008 &  0.007 &  0.008 \\
010005 & 05:53:01.69 & +32:43:25.9 & 15.816 & 15.076 & 14.715 &  0.008 &  0.007 &  0.008 \\
010006 & 05:53:07.01 & +32:43:16.1 & 15.605 & 15.096 & 14.825 &  0.008 &  0.007 &  0.008 \\
010007 & 05:53:10.26 & +32:44:20.4 & 16.564 & 15.396 & 14.770 &  0.008 &  0.007 &  0.008 \\
010008 & 05:53:04.83 & +32:45:00.3 & 15.902 & 15.399 & 15.117 &  0.008 &  0.007 &  0.008 \\
010009 & 05:52:55.02 & +32:45:35.6 & 16.223 & 15.457 & 15.062 &  0.008 &  0.007 &  0.008 \\
010010 & 05:53:10.44 & +32:44:22.1 & 16.184 & 15.629 & 15.329 &  0.008 &  0.007 &  0.008 \\
\enddata
\tablenotetext{a}{The complete version of this table is in the electronic edition of the Journal. The printed edition contains only a sample.}
\tablenotetext{b}{The first two digits in the ID give the mosaic chip number on which the star was detected.}
\label{tab:phot_catalog}
\end{deluxetable}

\clearpage

\begin{deluxetable}{lrrrrl}
\rotate
\tabletypesize{\footnotesize}
\tablewidth{0pc}
\tablecaption{Previous Determinations of the Fundamental Parameters for M37}
\tablehead{\colhead{Study} & \colhead{$(m-M)_{V}$} & \colhead{$E(B-V)$} & \colhead{Metallicity} & \colhead{Age (Myr)} & \colhead{Isochrone Source}}
\startdata
\citet{Mermilliod.96} & 11.50 & 0.29 & $Z = 0.02$ & 450 & \citet{Bertelli.94} \\
\citet{Kalirai.01} & $11.65 \pm 0.13$ & $0.23 \pm 0.03$ & $Z = 0.02$ & 520 & \citet{Ventura.98} \\
\citet{Kiss.01} & $11.48 \pm 0.13$ & $0.29 \pm 0.03$ & $Z = 0.02$ & 450 & \citet{Bertelli.94} \\
\citet{Nilakshi.02} & $11.6 \pm 0.15$ & $0.30 \pm 0.04$ & $Z = 0.008$ & 400 & \citet{Girardi.00} \\
\citet{Sarajedini.04} & $11.57 \pm 0.16$ & $0.27 \pm 0.03$ & $[Fe/H] = +0.09$ & $\cdots$ & \\
\citet{Kalirai.05} & 11.50 & $0.23 \pm 0.01$ & $Z = 0.011 \pm 0.001$ & 650 & \citet{Ventura.98} \\
\citet{Kang.07} & 11.4 & 0.21 & $Z = 0.019$ & 450 & \citet{Girardi.00} \\
This Paper & $11.57 \pm 0.13$ & $0.227 \pm 0.038$ & $[Fe/H] = +0.045 \pm 0.044$ & $485$ \\
\enddata
\label{tab:sumprevisocresults}
\end{deluxetable}

\clearpage

\begin{deluxetable}{lrrrrrrr}
\tabletypesize{\scriptsize}
\tablewidth{0pc}
\tablecaption{Systematic Errors in Isochrone Fitting for M37}
\tablehead{\colhead{Source} & \colhead{$\Delta$Quantity} & \colhead{$\Delta$age (Myr)} & \colhead{$\Delta(m-M)_{V}$} & \colhead{$\Delta[M/H]$} & \colhead{$\Delta E(B-V)$} & \colhead{$\Delta E(V-I_{C})$} & \colhead{$\Delta E(V-K_{S})$}}
\startdata
Helium Abundance ($Y$) & $\pm 0.009$ & $\mp 5$ & $\mp 0.027$ & $\cdots$ & $\cdots$ & $\cdots$ & $\cdots$ \\
Calibration & $\cdots$ & $\cdots$ & $\pm 0.010$ & $\pm 0.010$ & $\pm 0.002$ & $\pm 0.003$ & $\pm 0.006$ \\
Membership Selection & $\cdots$ & $\pm 15$ & $\pm 0.065$ & $\pm 0.035$ & $\pm 0.012$ & $\pm 0.010$ & $\pm 0.026$ \\
Fitting Range & $\cdots$ & $\pm 18$ & $\pm 0.088$ & $\pm 0.008$ & $\pm 0.012$ & $\pm 0.012$ & $\pm 0.034$ \\
$\Delta B$ & $\pm 0.025$ & $\cdots$ & $\cdots$ & $\cdots$ & $\pm 0.025$ & $\cdots$ & $\cdots$ \\
$\Delta V$ & $\pm 0.021$ & $\cdots$ & $\pm 0.021$ & $\cdots$ & $\mp 0.021$ & $\pm 0.021$ & $\pm 0.021$ \\
$\Delta I_{C}$ & $\pm 0.05$ & $\cdots$ & $\cdots$ & $\cdots$ & $\cdots$ & $\mp 0.05$ & $\cdots$ \\
$\Delta K_{S}$ & $\pm 0.007$ & $\cdots$ & $\cdots$ & $\cdots$ & $\cdots$ & $\cdots$ & $\mp 0.007$ \\
\hline
Total Systematic & $\cdots$ & $\pm 24$ & $\pm 0.12$ & $\pm 0.037$ & $\pm 0.037$ & $\pm 0.057$ & $\pm 0.049$ \\
Total Internal & $\cdots$ & $\pm 15$ & $\pm 0.046$ & $\pm 0.024$ & $\pm 0.008$ & $\pm 0.008$ & $\pm 0.018$ \\
Total & $\cdots$ & $\pm 28$ & $\pm 0.13$ & $\pm 0.044$ & $\pm 0.038$ & $\pm 0.058$ & $\pm 0.052$ \\
\enddata
\label{tab:systematics}
\end{deluxetable}

\clearpage

\begin{deluxetable}{lrr}
\tabletypesize{\footnotesize}
\tablewidth{0pc}
\tablecaption{Parameters for the Open Cluster M37 determined in this paper.}
\tablehead{\colhead{Parameter [Units]} & \colhead{Value} & \colhead{Error}}
\startdata
Age\tablenotemark{a} [Myr] & 485 & $\pm 28$ \\
Distance [pc] & 1490 & $\pm 120$ \\
$[M/H]$ [dex] & $+0.045$ & $\pm 0.044$ \\
$(m-M)_{V}$ [mag] & 11.572 & $\pm 0.13$ \\
$E(B-V)$ [mag] & 0.227 & $\pm 0.038$ \\
$E(V-I_{C})$ [mag] & 0.355 & $\pm 0.058$ \\
$E(V-K_{S})$ [mag] & 0.695 & $\pm 0.052$ \\
Center RA\tablenotemark{b} (J2000) [h:m:s] & 05:52:17.6 & $\pm 30\arcsec$ \\
Center Dec\tablenotemark{b} (J2000) [d:m:s] & +32:32:08 & $\pm 30\arcsec$ \\
Tidal Radius [$\arcmin$] & 50 & $\pm 15$ \\
Core Radius ($1.5 M_{\odot} < M < 2.6 M_{\odot})$ [$\arcmin$] & 4.0 & $\pm 0.4$ \\
Core Radius ($1.13 M_{\odot} < M < 1.5 M_{\odot})$ [$\arcmin$] & 6.1 & $\pm 0.6$ \\
Core Radius ($0.77 M_{\odot} < M < 1.13 M_{\odot})$ [$\arcmin$] & 6.3 & $\pm 0.6$ \\
Core Radius ($0.4 M_{\odot} < M < 0.77 M_{\odot})$ [$\arcmin$] & 6.6 & $\pm 0.5$ \\
Total Number of Stars & 4840 & $\pm 500$ \\
Total Mass [$M_{\odot}$] & 3640 & $\pm 170$ \\
Systemic RV of dwarf stars [km/s] & 9.4 & $\pm 0.2$ \\
\enddata
\tablenotetext{a}{Assuming an age for the Hyades of 550 Myr without overshooting. With overshooting, the age of M37 would be $550 \pm 30$ Myr.}
\tablenotetext{b}{Value from \citet{Kalirai.01}}
\label{tab:M37finalparams}
\end{deluxetable}

\clearpage

\begin{deluxetable}{rrr}
\tabletypesize{\footnotesize}
\tablewidth{0pc}
\tablecaption{Mass Function From $gri$ Photometry}
\tablehead{\colhead{Mass ($M_{\odot}$)} & \colhead{Number per Unit Mass\tablenotemark{a}} & \colhead{Error}}
\startdata
0.35 & 3574.92 & 1987.08 \\
0.45 & 5192.01 &  512.35 \\
0.55 & 5347.91 &  539.10 \\
0.65 & 4001.13 &  435.23 \\
0.75 & 4015.29 &  410.85 \\
0.85 & 4275.32 &  453.64 \\
0.95 & 3238.74 &  396.96 \\
1.05 & 2582.13 &  323.80 \\
1.15 & 1539.49 &  255.77 \\
1.25 & 1104.70 &  229.65 \\
1.35 &  867.67 &  154.80 \\
\enddata
\tablenotetext{a}{The mass bin width used is $0.1 M_{\odot}$}
\label{tab:M37_MF}
\end{deluxetable}

\clearpage

\begin{deluxetable}{rrr}
\tabletypesize{\footnotesize}
\tablewidth{0pc}
\tablecaption{Mass Function From 2MASS Photometry}
\tablehead{\colhead{Mass ($M_{\odot}$)} & \colhead{Number per Unit Mass\tablenotemark{a}} & \colhead{Error}}
\startdata
1.10 & 1530.01 &  484.45 \\
1.30 &  666.35 &  324.68 \\
1.50 &  931.36 &  222.42 \\
1.70 &  710.00 &  141.12 \\
1.90 &  348.18 &   93.85 \\
2.10 &  400.45 &   85.99 \\
2.30 &  222.73 &   69.52 \\
2.50 &  245.45 &   67.63 \\
\enddata
\tablenotetext{a}{The mass bin width used is $0.2 M_{\odot}$}
\label{tab:M37_MF_2MASS}
\end{deluxetable}

\clearpage

\clearpage

\clearpage

\clearpage

\clearpage

\clearpage

\clearpage

\clearpage

\clearpage

\clearpage

\clearpage

\clearpage

\clearpage

\clearpage

\clearpage

\clearpage

\clearpage

\clearpage

\clearpage

\clearpage

\clearpage

\end{document}